\documentclass{aa}
\usepackage{amssymb}
\usepackage{multirow}
\usepackage{rotating}
\usepackage{ctable}
\usepackage{morefloats}
\usepackage{graphicx,epsfig,fancyhdr,psfig,rotating,amsmath,epsf,txfonts,natbib,epstopdf,multirow}
\usepackage[colorlinks, citecolor=blue]{hyperref}
\begin{document}

\title{Coronal hole boundaries at small scales: IV. SOT view}
\subtitle{Magnetic field properties of small-scale transient brightenings in coronal holes}

\author{Z. Huang\inst{1} \and M.~S. Madjarska\inst{1,2} \and J.~G. Doyle\inst{1} \and D.~A. Lamb\inst{3}}

\offprints{zhu@arm.ac.uk}
\institute{Armagh Observatory, College Hill, Armagh BT61 9DG, N. Ireland, UK
\and UCL-Mullard Space Science Laboratory, Holmbury St Mary, Dorking, Surrey, RH5 6NT, UK
\and Department of Space Studies, Southwest Research Institute, 1050 Walnut Street Suite 300, Boulder CO 80302, USA}

\date{Received date, accepted date}

\abstract
{{We study the magnetic properties of small-scale transients in coronal holes and a few in 
the quiet Sun identified in X-ray observations in paper~II and analysed in spectroscopic data in paper~III.}}
{We aim to investigate the role of small-scale transients in the evolution of the magnetic field in an equatorial coronal hole.}
{Two sets of observations of an equatorial coronal hole and another two in quiet-Sun regions were analysed using 
longitudinal magnetograms taken by the Solar Optical Telescope. An automatic feature tracking program, SWAMIS, was 
used to identify and track the magnetic features. Each event was then visually analysed in detail.}
{In both coronal holes and quiet-Sun regions, all brightening events are associated with bipolar regions and 
caused by magnetic flux emergence followed by cancellation with the pre-existing and/or newly emerging magnetic 
flux. In the coronal hole, 19 of 22 events have a single stable polarity which does not change its position in time. 
In eleven cases this is the dominant polarity. 
The dominant flux of the coronal hole form the largest concentration of magnetic flux in terms of size while the opposite polarity is
distributed in small concentrations. We found that in the coronal hole the number of magnetic elements of the dominant polarity 
is four times higher than the non-dominant one. The supergranulation configuration appears to preserve 
its general shape during approximately nine hours of observations although the large concentrations (the dominant polarity) in the network  
did evolve and/or were slightly displaced, and their 
strength either increased or decreased.
The emission fluctuations/radiance oscillations seen in the X-ray bright points are associated with reoccurring magnetic cancellation
in the footpoints. {Unique observations of an X-ray jet reveal similar magnetic behaviour in the footpoints, i.e. cancellation of the opposite polarity magnetic flux. We found that the magnetic flux 
cancellation rate during the jet  is much higher than in bright points.} Not all magnetic cancellations result in an X-ray enhancement, 
suggesting that there is a threshold of the 
amount of magnetic flux involved in a cancellation above which brightening would occur at X-ray temperatures.}
{Our study demonstrates that the magnetic flux in coronal holes 
is continuously `recycled' through magnetic reconnection which is responsible for the formation of numerous 
small-scale transient events. The open magnetic flux forming the coronal-hole phenomenon is largely involved in 
these transient features. The question on whether this open flux is transported as a result of the formation 
and evolution of these transient events, however, still remains open.}

\authorrunning{Huang Z. et al.}
\titlerunning{Coronal hole boundaries at small scales: IV}

\maketitle

\section{Introduction}
\label{sect:intro}
Coronal holes (CHs) are regions on the Sun where the emission of coronal lines is significantly reduced. CHs were found
to be the main source regions of the fast solar wind \citep{1973SoPh...29..505K} while coronal hole boundaries are 
believed to be the regions where the slow solar wind originates. We first studied the small-scale evolution of CHs 
and their boundaries using spectroscopic observations from SUMER \citep{2004ApJ...603L..57M}. This was followed by three 
studies. The first one investigated dynamic phenomena at the coronal hole boundaries (CHBs) using 
TRACE (The Transition Region And Coronal Explorer) and EIT (Extreme-ultraviolet Imaging Telescope) onboard SoHO 
\citep[][hereafter paper I]{2009A&A...503..991M}. The second study by \citet[][hereafter paper II]{2010A&A...516A..50S} 
automatically identified X-ray transient brightenings in CHs and the quiet-Sun regions in observations from XRT 
(The X-Ray Telescope) onboard the Hinode satellite. Next, \citet[][hereafter paper III]{madj2012} analysed the plasma 
properties of all the events which were identified in paper II having simultaneous spectral observations taken with the EIS 
(Extreme-ultraviolet Imaging Spectrometer) and SUMER (Solar Ultraviolet Measurements of Emitted Radiation) instruments 
onboard Hinode and SoHO, respectively. By studying tens of events, the authors found that events in the CHs and quiet-Sun regions reached 
similar temperatures and electron densities, but events in CHs and their boundaries presented higher dynamics than 
events in the quiet Sun. Background information on coronal holes and an overview on the results of the papers published 
earlier are given in \citet[][and the references therein]{madj2012}. These studies showed that the brightening events, 
including coronal bright points (BPs) and jets, are associated with the small-scale dynamics of CHBs.

\begin{table*}
\centering
\caption{Description of the observations used in this study.}
\begin{tabular}{l c c c c c}
\hline\hline
Date &Observing period&\multicolumn{2}{c}{Cadence (seconds)}&Observed&SOT FOV size\\
&(UT)&XRT observations&SOT magnetograms&region&\\

\hline
2007-Nov-09&06:38 $\rightarrow$ 14:59&40&90&Coronal hole&261\arcsec$\times$148\arcsec\\
2007-Nov-12&01:21 $\rightarrow$ 10:57&40&90&Coronal hole&261\arcsec$\times$148\arcsec\\
2009-Jan-10&11:30 $\rightarrow$ 16:59&60&45&Quiet Sun&46\arcsec$\times$148\arcsec\\
2009-Jan-13&11:22 $\rightarrow$ 17:34&60&45&Quiet Sun&46\arcsec$\times$148\arcsec\\
\hline
\end{tabular}
\label{tab:obs}
\end{table*}


\begin{figure*}[!htb]
 \centering
 \includegraphics[width=8cm,trim=5mm 15mm 28mm 28mm,clip]{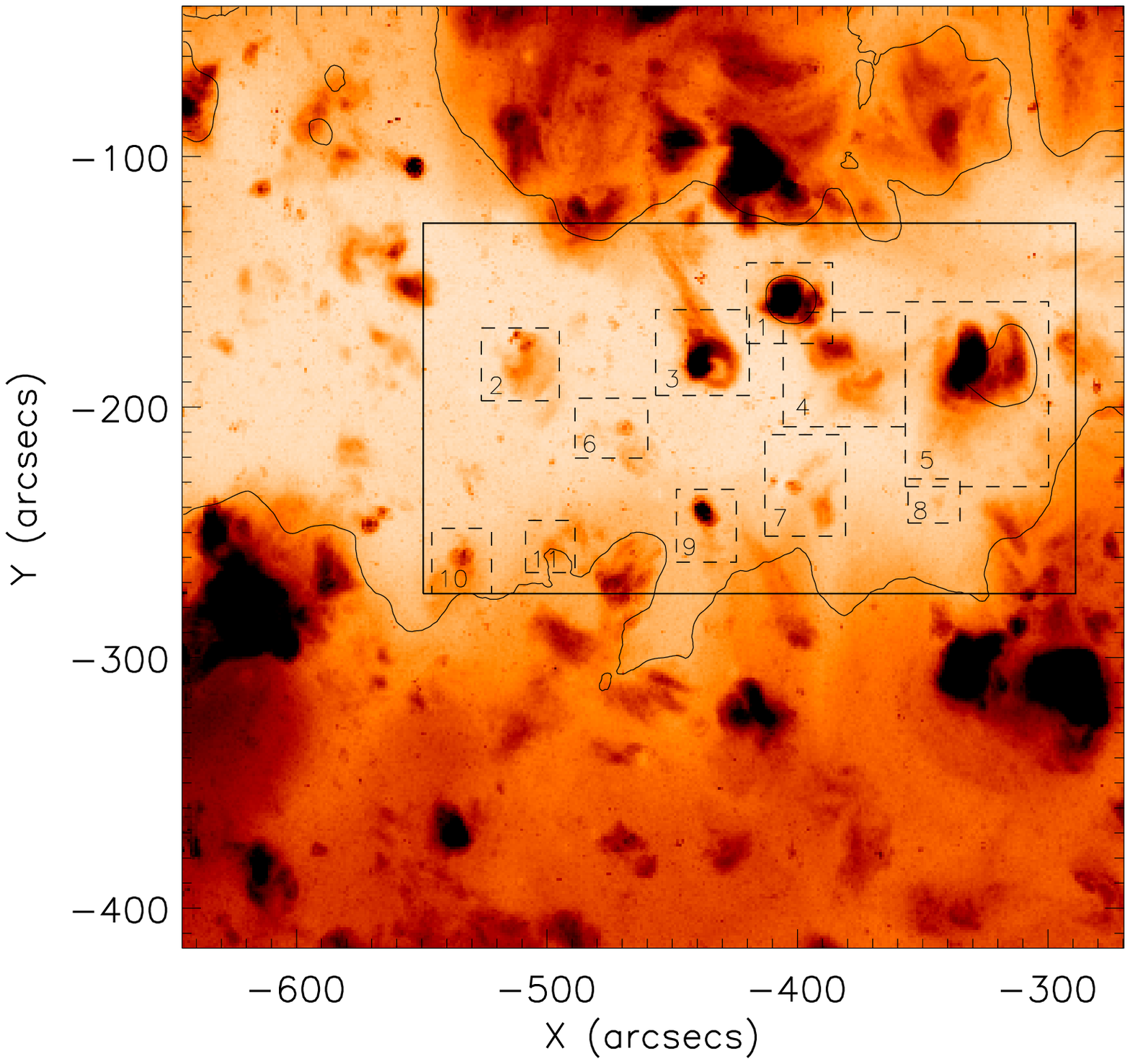}\includegraphics[width=8cm,trim=5mm 15mm 28mm 28mm,clip]{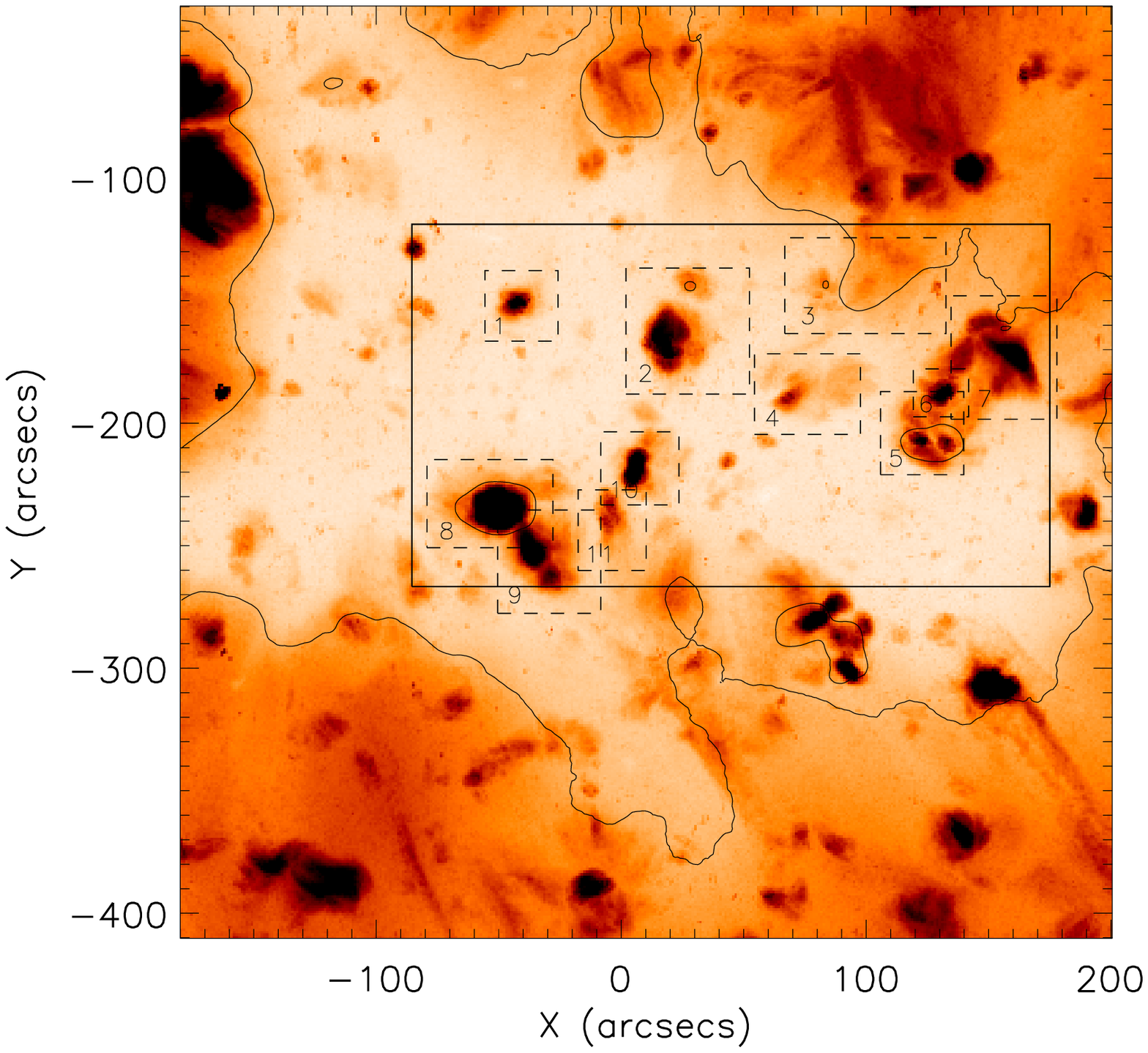}
 \includegraphics[width=8cm,trim=5mm 15mm 28mm 28mm,clip]{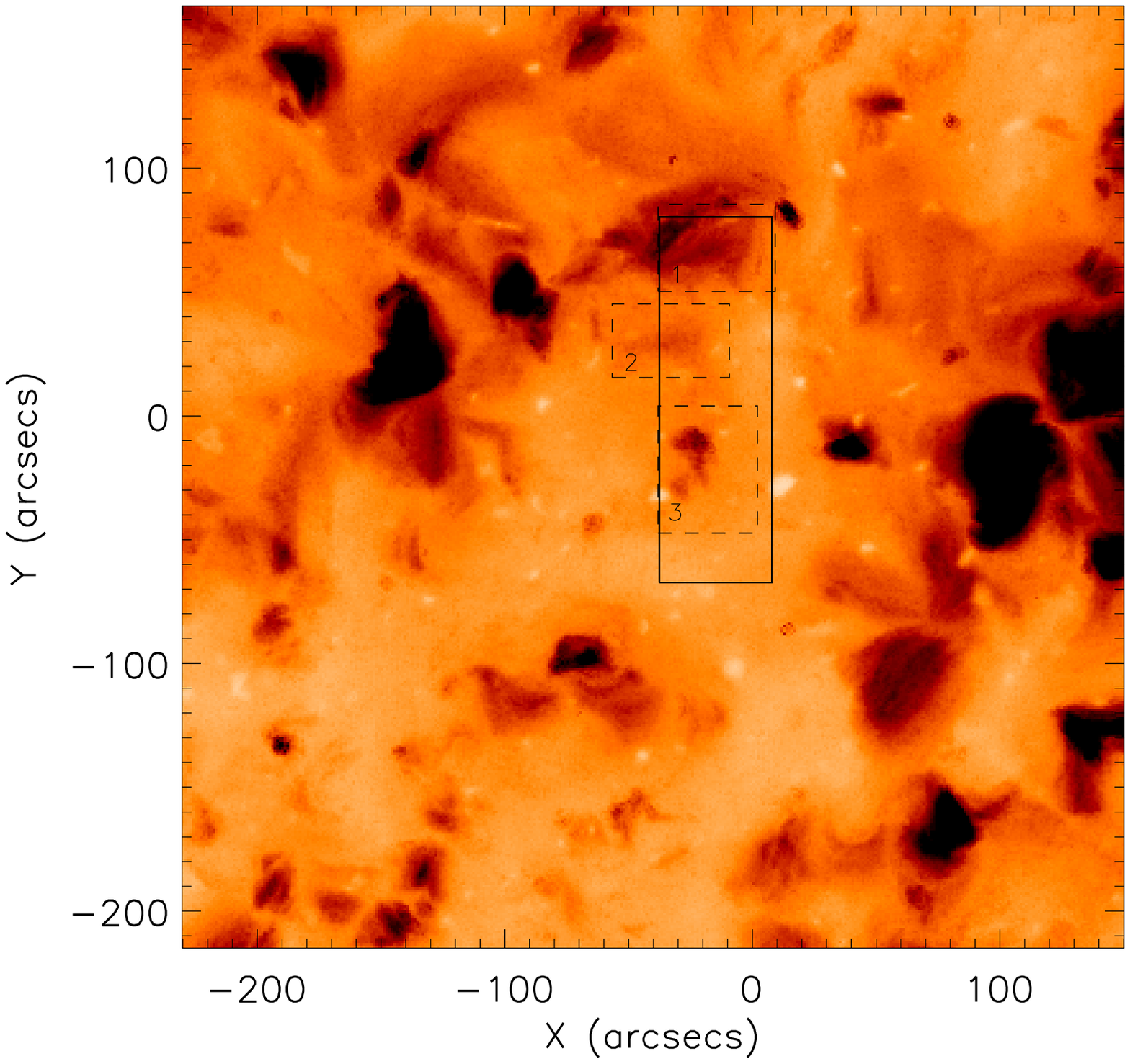}\includegraphics[width=8cm,trim=5mm 15mm 28mm 28mm,clip]{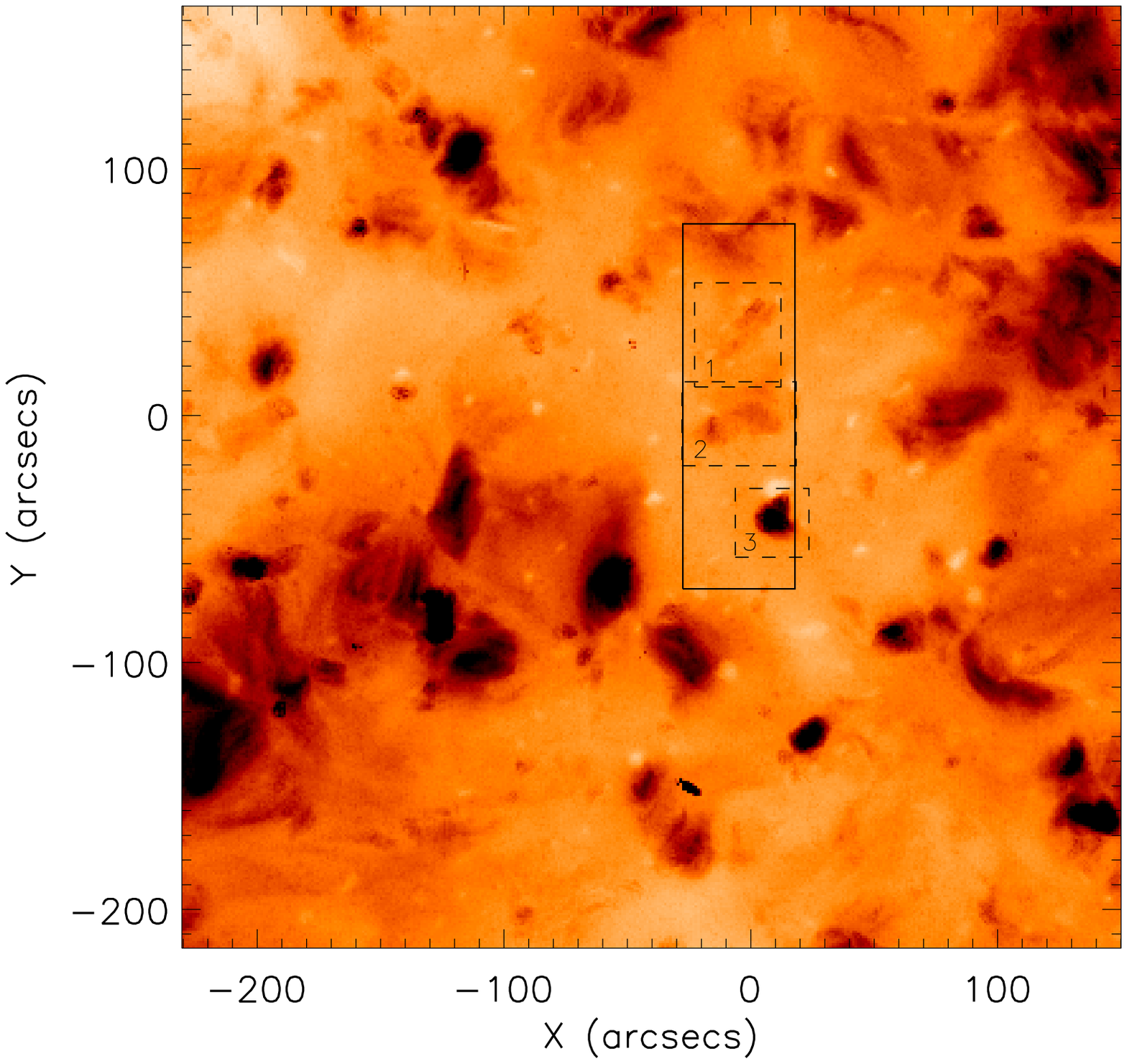}
 \caption{Generated X-ray images (see the text in Section~2.2) of an equatorial CH on 2007 November 9 (top left) and 
 November 12 (top right), the quiet Sun 2009 January 10 (bottom left) and January 13 (bottom right). The 
 images are shown with an inverted colour table. The SOT/Hinode field-of-views are marked as rectangles~(straight solid lines). Features 
 within the dashed-line boxes are the events studied in this paper. For the two CH datasets shown in the top 
 panel, the black solid contour plot outlines the coronal hole boundary as seen at the beginning of each observing period.}
 \label{fig1}
\end{figure*}

\begin{figure}[!ht]
 \centering
 \resizebox{\hsize}{!}{\includegraphics[trim=2mm 2mm 25mm 5mm,clip]{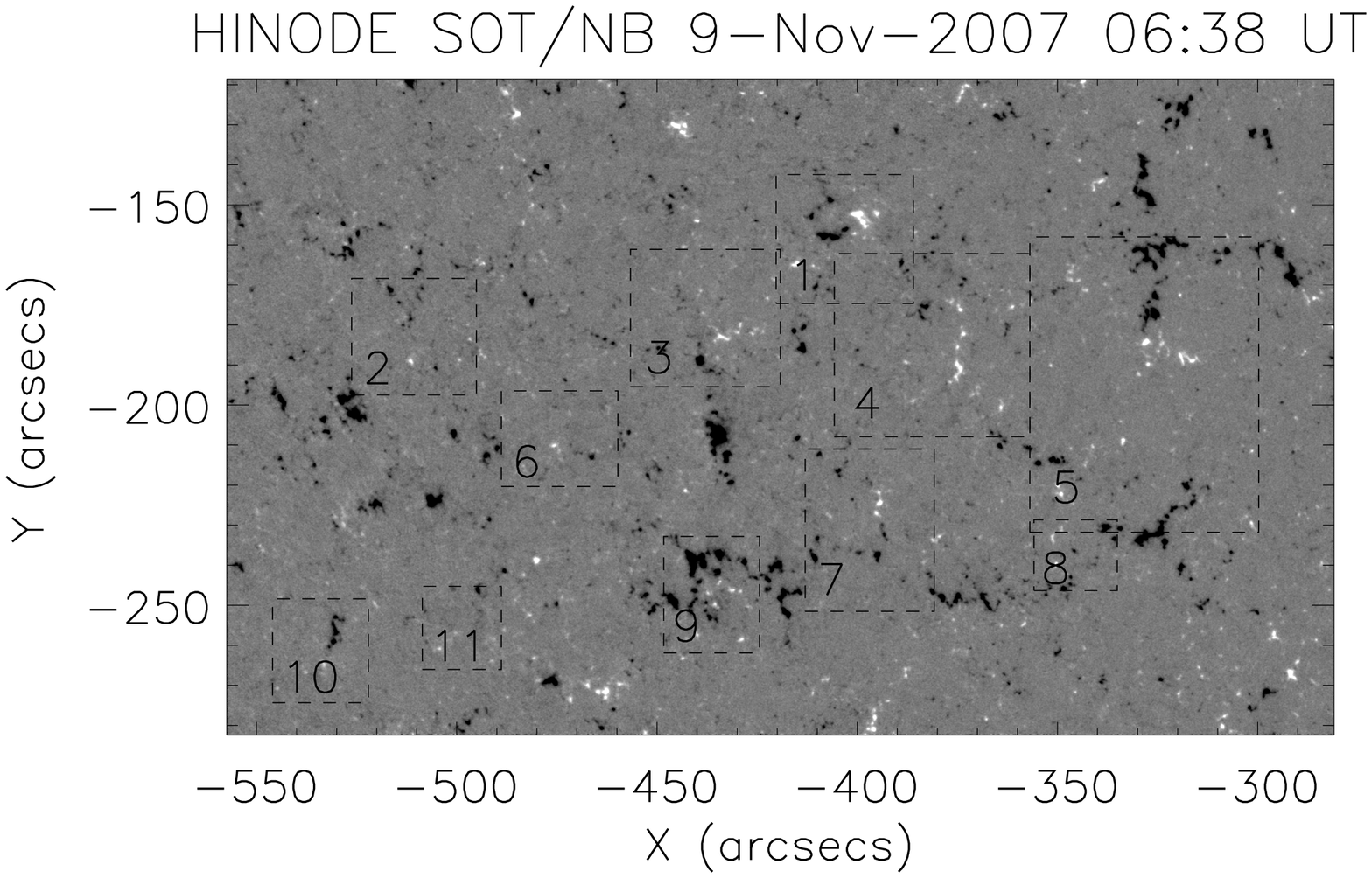}}
 \resizebox{\hsize}{!}{\includegraphics[trim=2mm 2mm 25mm 5mm,clip]{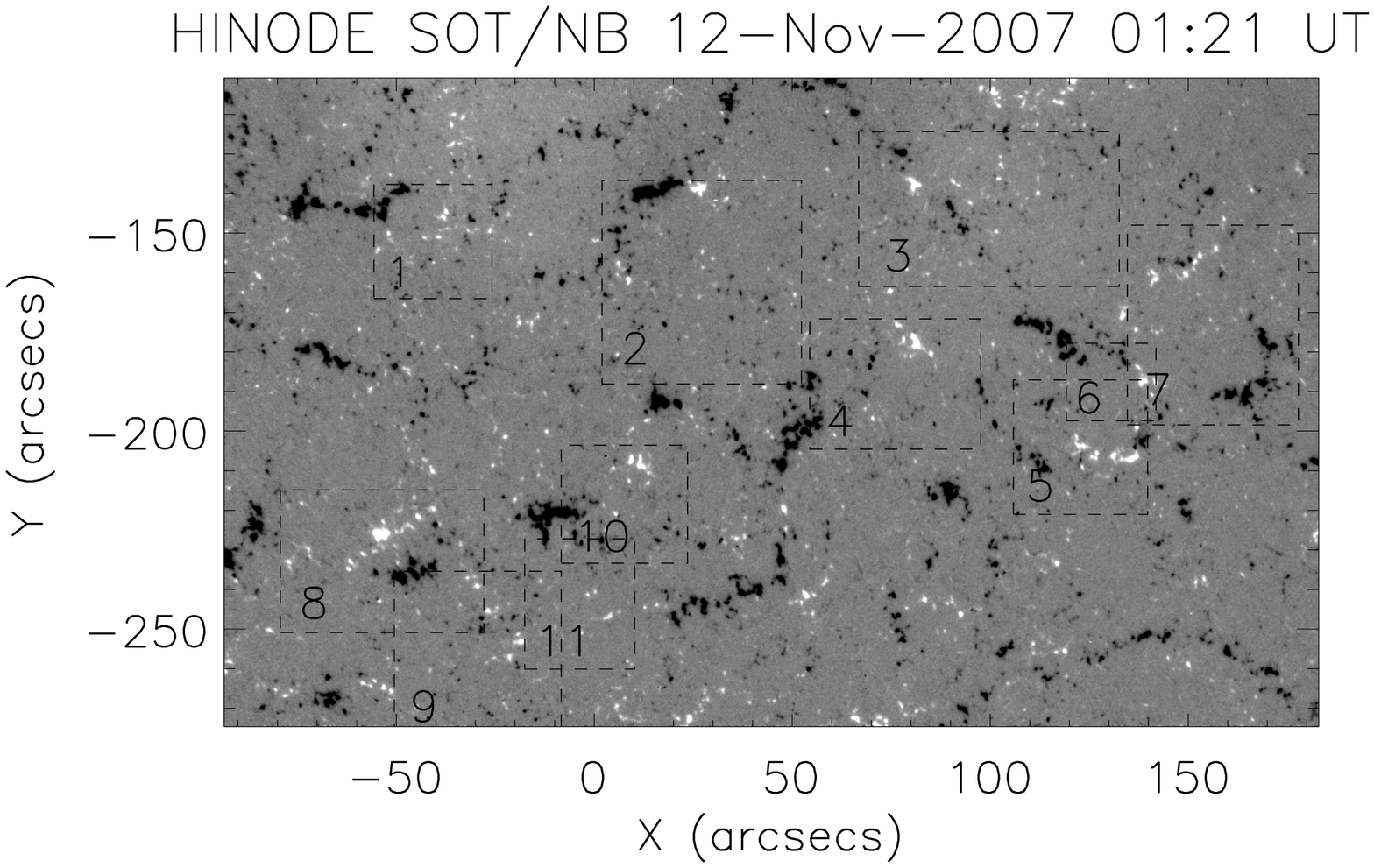}}
 \caption{Full field-of-view of the SOT longitudinal magnetograms of the coronal hole analysed in this study. Overplotted with dashed lines are the FOVs of the events identified in X-rays, which are shown in the top panels of Fig.\,\ref{fig1}.}
 \label{fig2}
\end{figure}

\begin{figure}[!ht]
 \centering
 \resizebox{\hsize}{!}{\includegraphics[trim=0.5cm 1cm 1.5cm 1.5cm,clip]{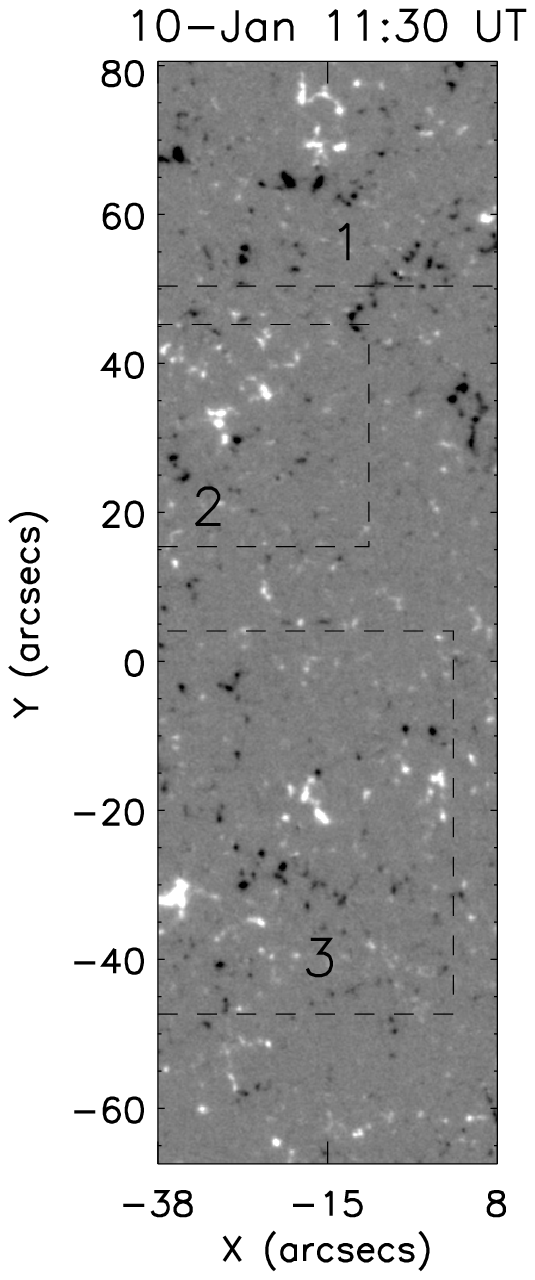}
			 \includegraphics[trim=0.5cm 1cm 1.5cm 1.5cm,clip]{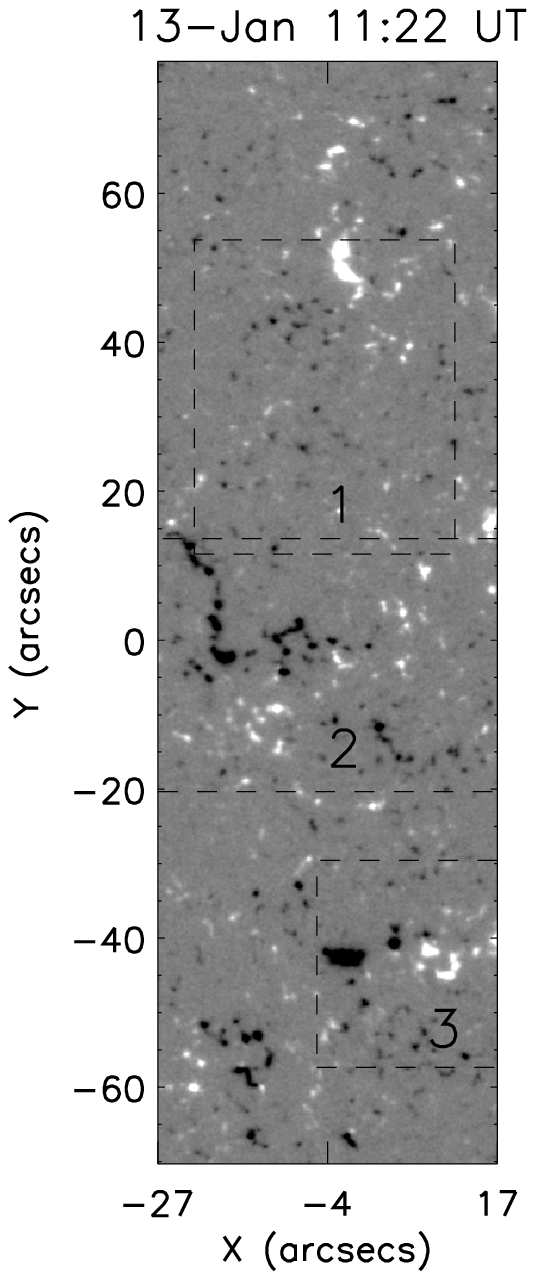}}
 \caption{Full field-of-view of the SOT longitudinal magnetograms of the quiet Sun analysed in this study. Overplotted with dashed lines are the FOVs of the events identified in X-rays, which are shown in the bottom panels of Fig.\,\ref{fig1}.}
 \label{fig3}
\end{figure}
\par
Coronal BPs are omnipresent in both CHs and the quiet Sun. They were first identified in soft X-rays \citep{vaiana1970a} 
and analysed in detail during the \textit{Skylab} mission 
\citep[][and references therein]{1976SoPh...49...79G, 1976SoPh...50..311G, 1992AnGeo..10...34H}. They are small (on 
average 20\arcsec--30\arcsec) and short-lived (from a few minutes to a few tens of hours) emission enhanced 
structures observed in the solar corona \citep[][and references therein]{1993SoPh..144...15W}. They can also be observed 
in the EUV wavelength bands \citep[][etc.]{1981SoPh...69...77H, 2001SoPh..198..347Z}. BPs represent small-scale loop 
structures first seen in an unique sequence of high-resolution Naval Research Laboratory (NRL) Skylab spectroheliograms 
with a spatial resolution of 2\arcsec \citep{1979SoPh...63..119S}. \citet{1979SoPh...63..119S} found that BPs consist 
of two or three miniature loops (2\,500~km in diameter and 12\,000 km long) evolving on a time scale of 6~min. 
\citet{1990ApJ...352..333H} confirmed this result showing that simultaneously measured peaks of emission in six 
different lines (emitted from the chromosphere to the corona) were not always co-spatial, implying that the BPs are 
composed of small-scale loops at different temperatures. The fine structure of BPs consisting of numerous loops was 
later confirmed by the high-resolution TRACE and now by XRT/Hinode and AIA/SDO observations.

\par
Studies of BPs have shown variations in their radiance lightcurves 
\citep[][etc.]{1977ApJ...218..286M, 1979SoPh...63..119S, 1981SoPh...69...77H}. \citet{2003A&A...398..775M} found six 
minutes radiance variations in SUMER transition region observations of a coronal BP. \citet{2004A&A...418..313U} derived 
intensity oscillations with time scales of 420--650~s while later \citet{2004A&A...425.1083U} reported oscillations
with periods ranging from 600~s to 1100~s. These oscillations were interpreted as an indicator of global magnetic-acoustic 
modes of the closed magnetic structures associated with BPs. Although oscillations with periods of a few minutes is more 
commonly known, longer periods were also found by \citet{2008A&A...489..741T}. The reason for these oscillations is still 
under debate. \citet{1988ApJ...330..474P} suggested they are repetitive small-scale flares, i.e. micro-flares. \citet{2008A&A...489..741T} 
suggested magneto-acoustic waves and/or a recurrence of magnetic reconnection. \citet{2004A&A...425.1083U} found that there 
is a one to one relation between the magnetic flux in the bipolarity and the EIT 195~\AA\ coronal emission during the 
growing and decaying phase for the two BPs studied. The authors concluded that their results give further support to the 
idea that magnetic reconnection involving the interaction of two magnetic polarities as the more likely operating mechanism. 

\par
Regarding magnetic fields, BPs are found to be associated with small bipolar regions 
\citep{1971IAUS...43..397K, 1977SoPh...53..111G, 1993SoPh..144...15W, 2001SoPh..201..305B}. With the Big Bear Solar Observatory 
(BBSO) magnetograms, \citet{1993SoPh..144...15W} investigated the correspondence of evolving magnetic features of 25 X-ray BPs. 
They found 22 BPs associated with converging of the magnetic features, while 18 were associated with cancelling magnetic 
features. They also found the BPs were more likely associated with pre-existing magnetic features rather than newly emerging 
ones. From higher resolution observations during the \textit{Yohkoh} mission, \citet{1994xspy.conf...21H} further 
demonstrated that most BPs were associated with magnetic cancellation. \citet{1999ApJ...510L..73P} 
found that the time evolution of several BPs observed by EIT 195~\AA\ images was strongly correlated with 
the variation of magnetic flux as determined from MDI photospheric magnetograms. By studying a BP observed in 
EIT 195~\AA\ together with its MDI photospheric magnetograms, \citet{2003A&A...398..775M} confirmed such a 
correlation. They also found that the BP existed until the complete cancellation of one of the polarities from the corresponding 
bipolar region. \citet{2004A&A...418..313U} also observed a linear dependence between the EIT 195~\AA\ intensity 
flux from a BP and the total magnetic flux of a photospheric bipolar region. They found that the increase of the coronal 
emission associated with a BP was linked to the emergence of a new magnetic field and the disappearance of coronal 
emission was associated with the cancellation of one of the polarities. In one of the cases, the disappearance in EUV emission took 
place three to four hours before the full cancellation of the weakest polarity.
\begin{figure*}[!ht]
 \centering
\includegraphics[scale=0.48]{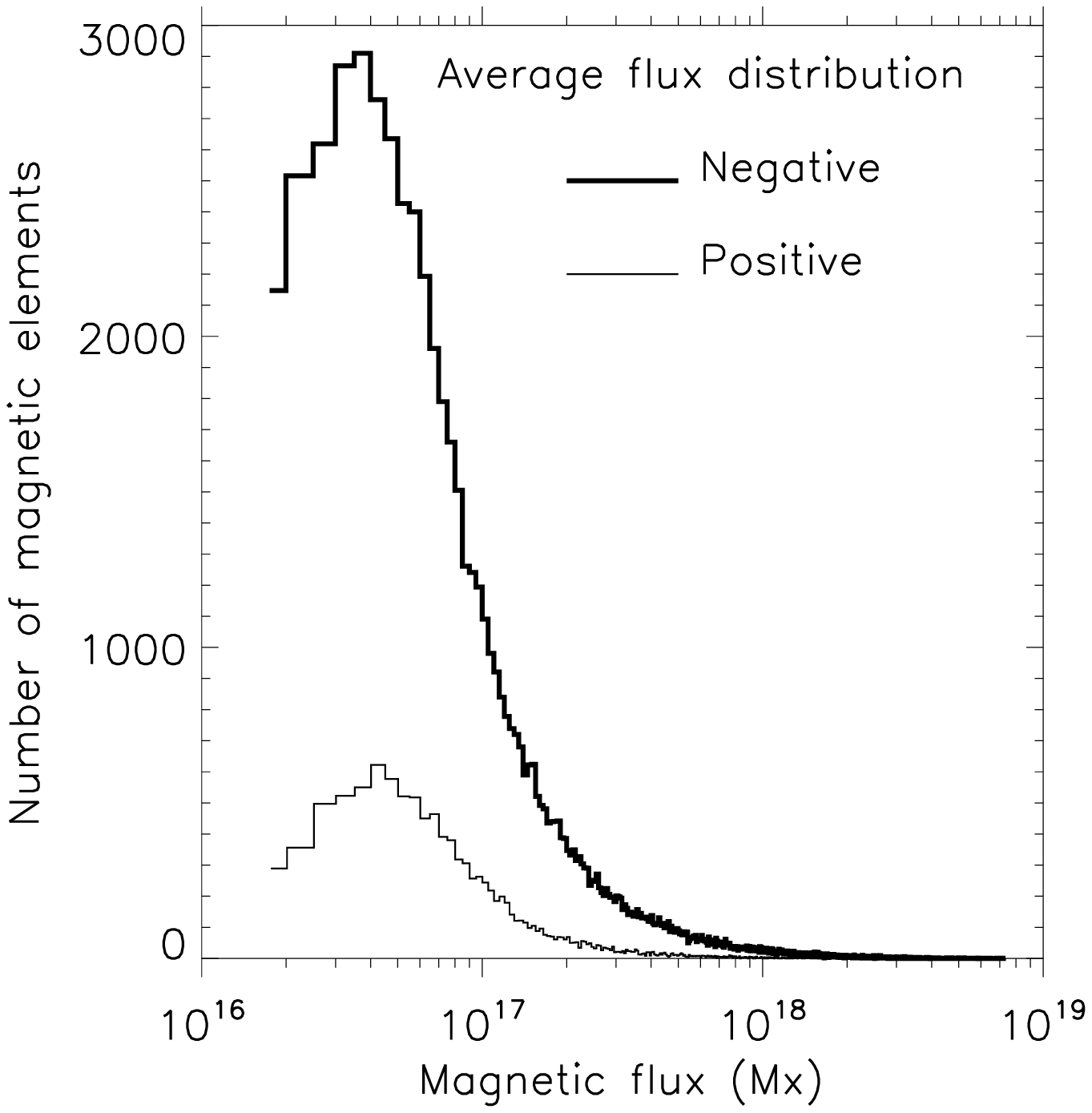}\includegraphics[scale=0.48]{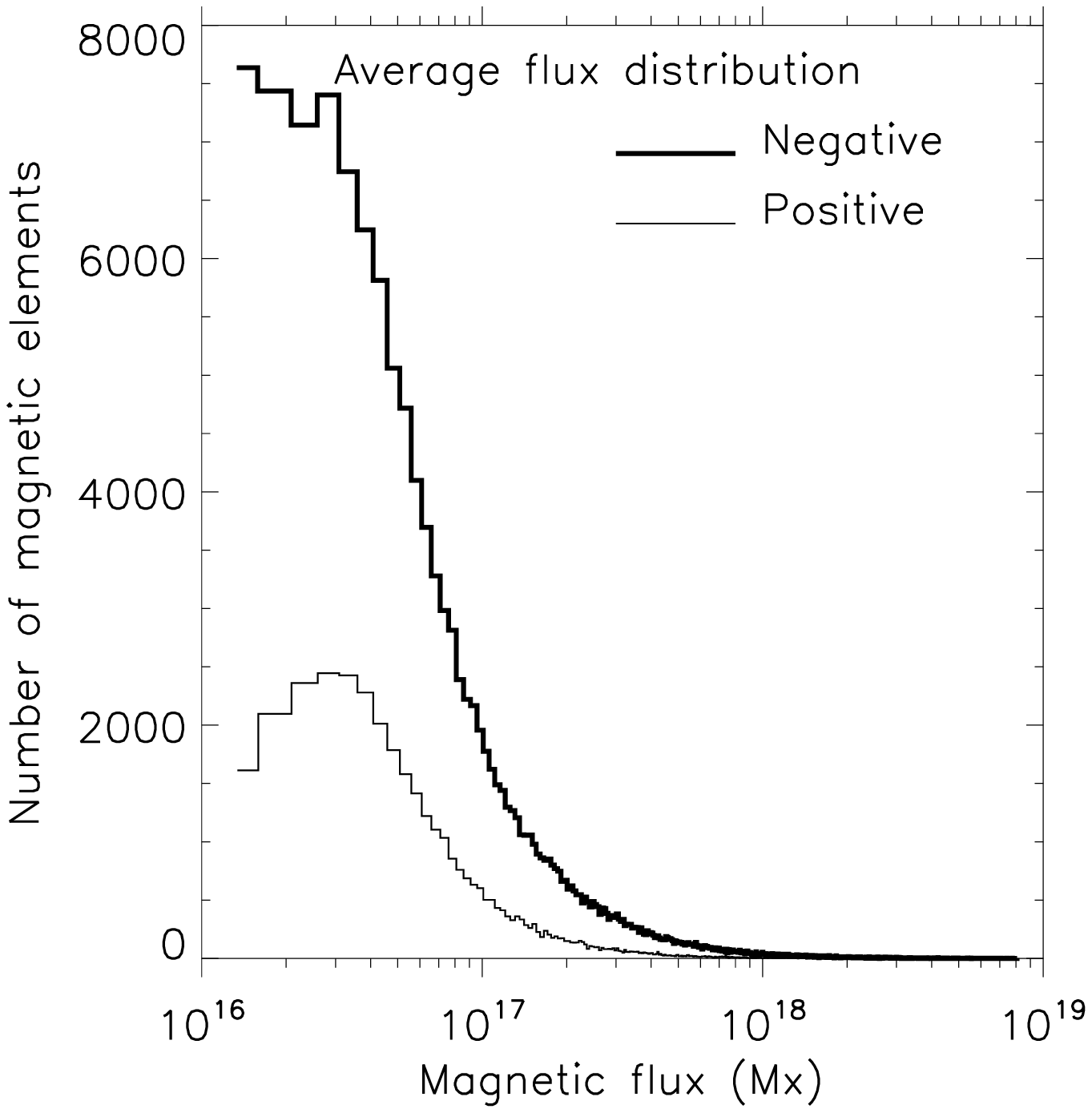}
 \includegraphics[scale=0.48]{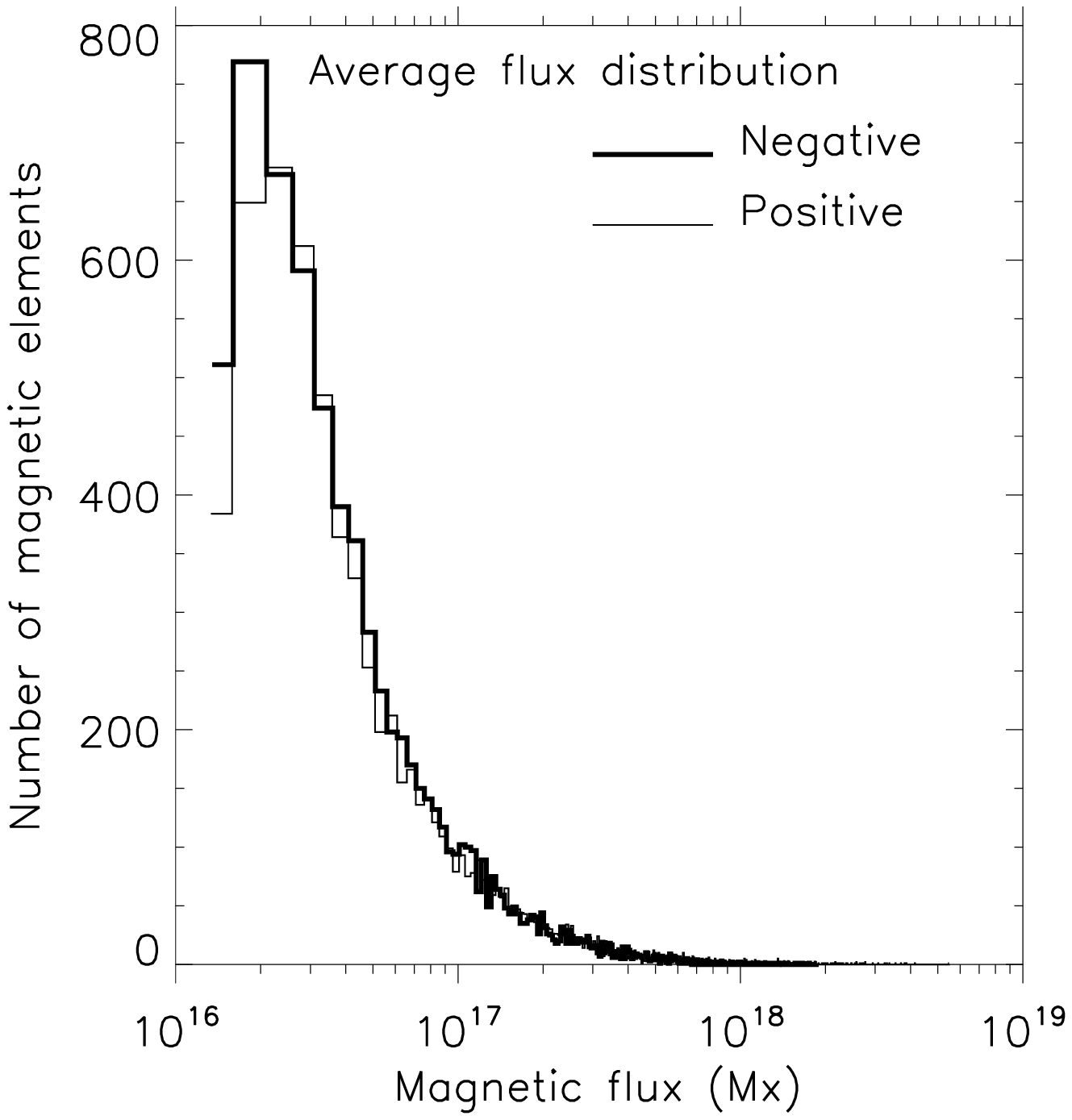}\includegraphics[scale=0.48]{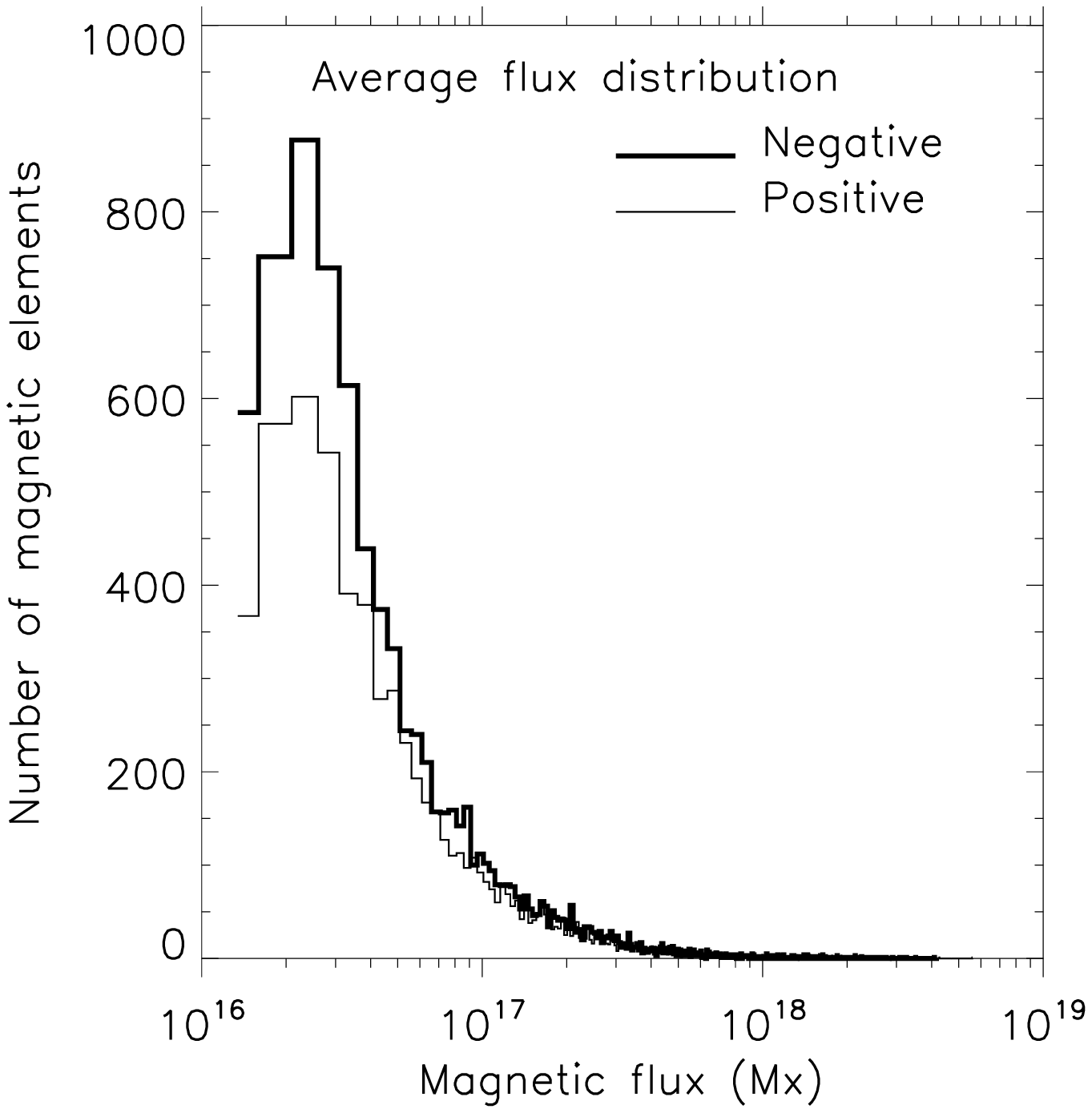}
 \caption{Magnetic flux distribution of the CH (top left November 9, right November 12) and quiet-Sun (bottom left January 10, right January 13) magnetic elements detected by SWAMIS.}
 \label{fig4}
\end{figure*}

\begin{figure}[!ht]
 \centering
 \resizebox{\hsize}{!}{\includegraphics{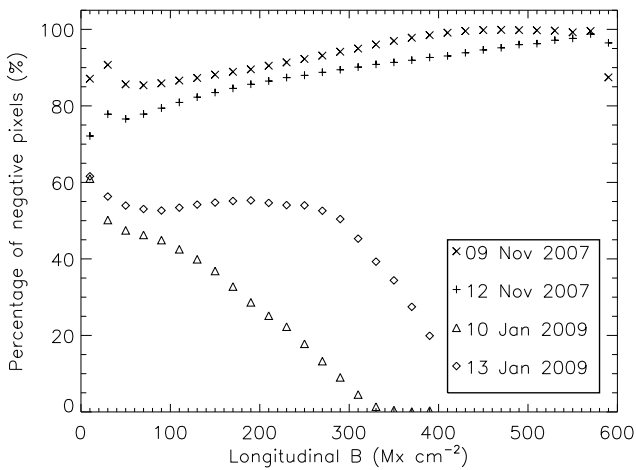}}
 \caption{Fractional number of pixels with negative longitudinal magnetic field for each day of the observations. }
 \label{fig5}
\end{figure}
\par

\par
As reviewed above, BPs are associated with magnetic cancellation which exhibits as magnetic flux 
disappearance in photospheric observations. What actually happens during the disappearance of the magnetic flux? 
\citet{1987ARA&A..25...83Z} suggested a few possibilities. One way is a simple submergence, 
when a pre-existing loop descends into the convection zone. Another scenario requires magnetic 
reconnection either above or below the photosphere, which is named reconnection submergence. The 
reconnection submergence was discussed in detail by \citet{1989ApJ...343..971V}. They pointed out that a 
certain distance between the two opposite polarities was required for this type of cancellation to happen. From simultaneous 
measurements of the magnetic field in the photosphere and chromosphere \citet{1999SoPh..190...35H} concluded that the magnetic 
flux ``is retracting below the surface for most, if not all, of the cancellation sites studied''. Furthermore, 
\citet{1994ApJ...427..459P} developed a model of BPs based on converging motions of magnetic features which 
can trigger magnetic reconnection and thereby energize a BP. The model was further developed in three-dimensions 
by \citet{1994SoPh..153..217P} and tested by two-dimensional numerical experiments by 
\citet{2006MNRAS.366..125V,2006MNRAS.369...43V}. In the model of \citet{1994ApJ...427..459P}, (called Converging Flux 
Model), a certain interaction distance is also required to start the reconnection and then energize the BP. Evidence of this model was provided by 
\citet{2003A&A...398..775M} who found that the BP appeared in EUV only 
when two opposite polarities were 10\arcsec\ apart. A possible evidence of magnetic reconnection in BPs was also 
confirmed by magnetic field reconfigurations \citep{2008A&A...492..575P, 2011A&A...526A.134A, 2012ApJ...746...19Z}.
\begin{figure*}[!ht]
 \centering
 \includegraphics[width=16cm, trim=42mm -15mm 0mm 60mm,clip]{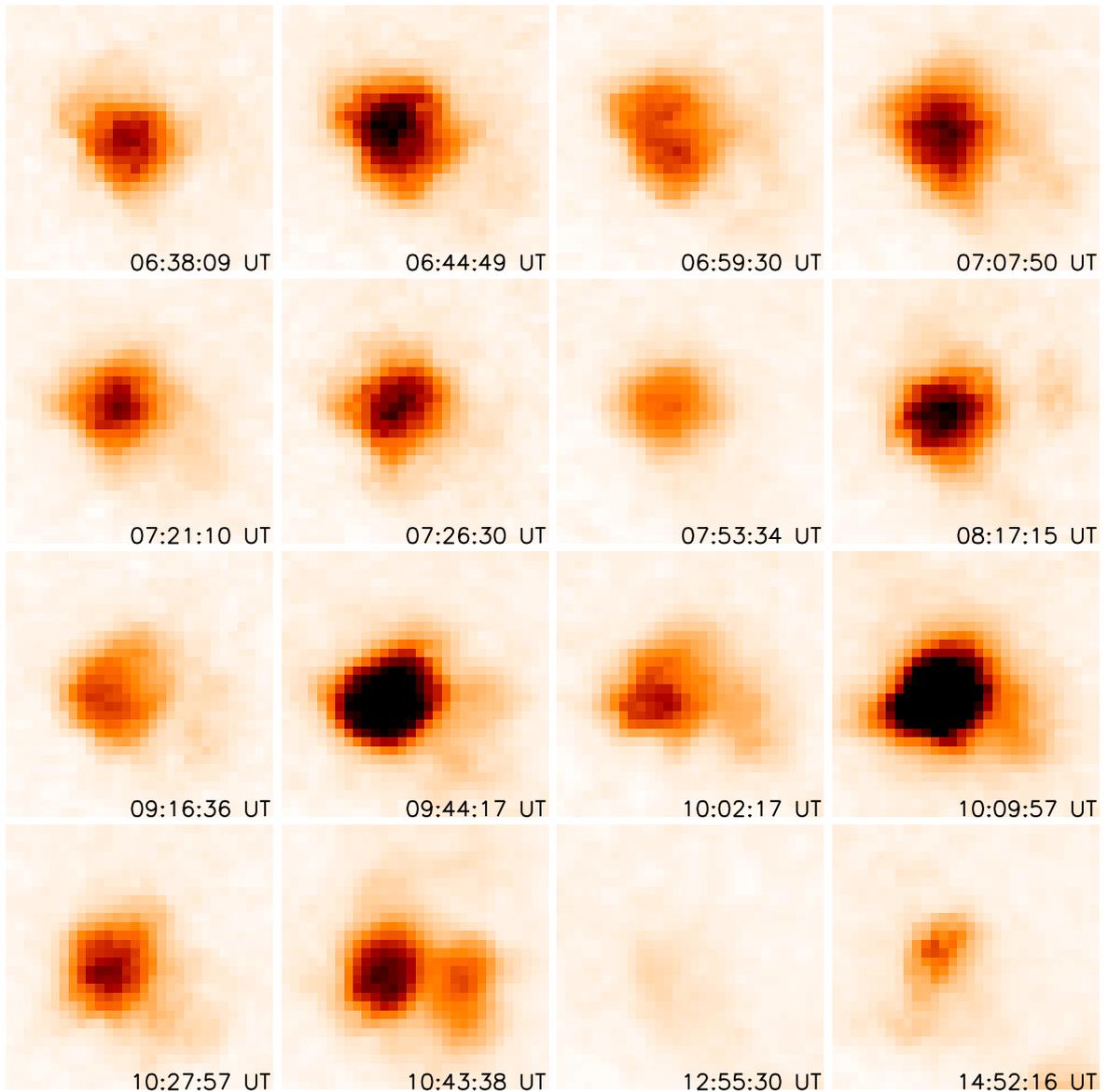}
 \caption{Temporal evolution of a CH bright point seen in X-ray on November 9 (event No. 1). The images are displayed in reversed colour table. The field-of-view size is 
 30\arcsec$\times$30\arcsec.  An animated version is available online, c.f. Fig.\,A\ref{fig20}.}
 \label{fig6}
\end{figure*}

\begin{figure*}[!ht]
 \centering
 \includegraphics[width=16cm, trim=42mm -15mm 0mm 60mm,clip]{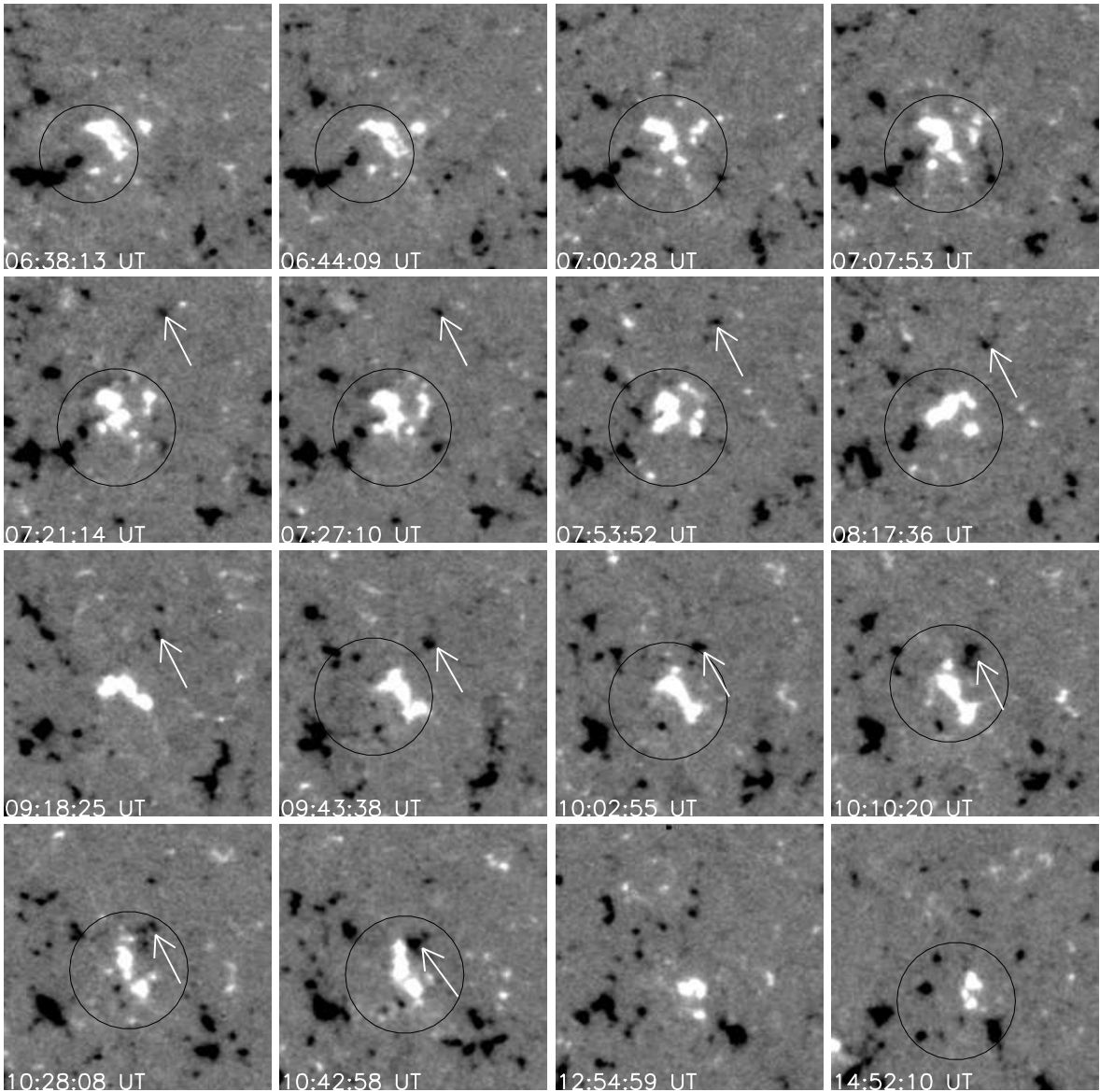}
 \caption{Longitudinal magnetic field images of the CH bright point shown in 
 Fig.\,\ref{fig6}. The field of view is the same as in Fig.\,\ref{fig6}, and the images are 
 scaled from --70~Mx~cm$^{-2}$ to +70~Mx~cm$^{-2}$. Whenever the magnetic cancellations are observed, the site is outlined by a black circle on the corresponding image. The circle is adjusted as precisely as possible to include only the involved magnetic features. A white arrow points at an emerging magnetic element. The field-of-view has a size of
 30\arcsec$\times$30\arcsec. An animated version is available online, c.f. Fig.\,A\ref{fig20}.}
 \label{fig7}
\end{figure*}

\par
X-ray jets are dynamic phenomena which represent collimated plasma flows from coronal BPs 
\citep[see paper II, III,][]{1998ApJ...508..899W, 2010ApJ...710.1806D}. They were first discovered in soft X-rays 
by \citet{1992PASJ...44L.173S} and are mostly associated with mixed polarity regions \citep{1998SoPh..178..379S}. Magnetic 
reconnection is believed to be the main mechanism of their formation \citep{1995Natur.375...42Y,2008ApJ...673L.211M}. 
A more detailed introduction on X-ray jets can be seen in paper~III and the references therein.

\par
As a continuation of papers I, II, and III and based on the brightenings identifications in paper II, 
 we focus here on the emergence, evolution and disappearance of magnetic flux associated with transient brightenings 
responsible for the small-scale evolution of coronal holes. A comparison with only a few events in the quiet Sun is also provided.
Such a detailed study of the magnetic field of small-scale transients in the solar atmosphere is unprecedented thanks to the 
 high-sensitivity, cadence and resolution of the SOT/Hinode data which has been pivotal for achieving the aim of our study. 
In Section \ref{sect:obs}, we 
describe the observations used in this study and the methods used in the data calibration and analysis. In Section 
\ref{sect:res}, we present the results. Our discussion and conclusions are given in Section \ref{sect:concl}.

\section{Observations and analysis}
\label{sect:obs}
\subsection{Data and calibrations}
The observations used in this study were taken by XRT \citep[X-Ray Telescope,][]{2007SoPh..243...63G} and SOT \citep[Solar Optical Telescope,][]{2008SoPh..249..167T} 
onboard Hinode. They include four datasets. Two of them were taken in November 2007 pointing at an equatorial 
coronal hole, and another two were taken in January 2009 in a quiet-Sun region. In Table\,\ref{tab:obs} we list all details 
of the observations. 

The XRT observations were taken with the Al$\_$Poly filter whose temperature response function peaks around 
8$\times10^6$\,K. The pixel size of the X-ray images is 1\arcsec$\times$1\arcsec. The standard procedures 
from the solar software (SSW) were used for the data reduction. The observations were carefully examined and any 
frames which were seriously affected by cosmic rays or instrumental effects were then excluded. In addition to the 
jitter correction, the images were further cross-correlated for the removal of the residual jitter.

\par
The SOT observations used in this study include two types of data, series of Stokes V and I polarimetric images taken 
with the Narrowband Filter Imager (NFI) and Stokes I, Q, U, V polarimetric spectra obtained with the Spectro-polarimeter 
(SP). The NFI filtergrams (FG) V and I images were taken in the Na~{\sc i}~5896~\AA\ spectral line which measures 
the magnetic field in the chromosphere \citep{2008SoPh..249..233I}. The pixel size of the magnetograms is 
0.16\arcsec$\times$0.16\arcsec. Each SP scan took about 20 minutes to obtain. The SP images have a pixel size of 0.32\arcsec$\times$0.32\arcsec. 
The standard software available from SSW was used to calibrate the data. Some of the V/I magnetograms which were 
badly affected by cosmic rays were not included. The time series of Stokes V/I magnetograms were further corrected for 
the jitter effect via a cross-correlation method. After this correction a region of 50 pixels (i.e. 8\arcsec) 
at each edge of the original \mbox{field-of-view} was discarded.
\begin{figure}[!ht]
 \centering
 \resizebox{\hsize}{!}{\includegraphics{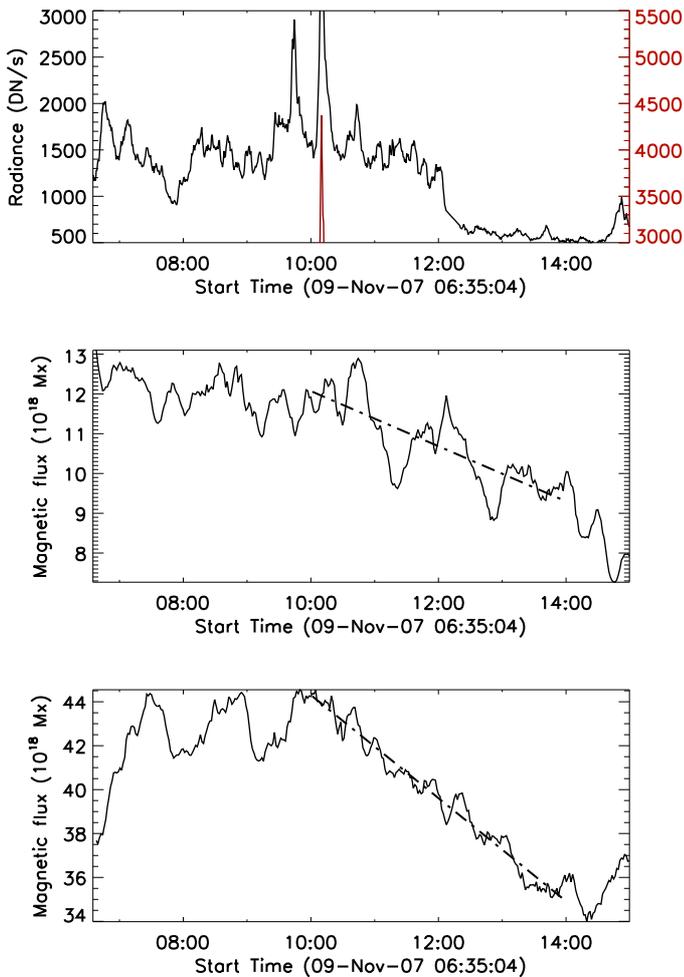}}
 \caption{Lightcurves~(top: X-ray radiance; middle: positive magnetic flux; bottom: negative magnetic flux) of a CH bright point 
 that is marked as No. 1 event on  November 9 in Fig.\,\ref{fig1}. The lightcurves are extracted from the FOV shown in 
 Figs.\,\ref{fig6}, and \ref{fig7}. The time~(x) axis is shown in unit of UT. {For better scaling the peak of the X-ray lightcurve is cut off and resumes from below 
 in red line. The dash-doted-lines in the middle and bottom panels are linear fits of the corresponding magnetic flux from 10:00\,UT to 14:00\,UT.}
 }
\label{fig8}
\end{figure}

\par
To convert the $V/I$ values into physical units, we adopted a method which was applied
by \citet{2007PASJ...59S.619C} to Fe~{\sc I}~ 6302.5~\AA. In this method, the magnetic strength ($B$) is assumed to be linear to $V/I$, 
i.e. $B=\beta\frac{V}{I}$. Although \citet{2007PASJ...59S.619C} 
mentioned that $\beta$ would vary with magnetic strength, they also suggested that it is a good approximation to assume it
as a constant in a region with similar solar activity. To obtain the best
linear fit, we used a slightly different formula, $B=\beta \frac{V}{I} + B_0$, where $\beta$ and $B_0$ 
are constants. We produced the longitudinal magnetogram from the SP scan with the standard method instructed by the 
\textit{SOT analysis guide}. We then averaged the FG V/I images obtained during this scan. Next, we re-binned the FG 
data to the same spatial scale as the SP magnetogram and co-aligned them. Finally, we applied a linear fit to the 
distribution of the SP longitudinal magnetic strength versus the FG $V/I$ values to get the calibration parameters 
of $\beta$ and $B_0$. The resulting values of the magnetic field can carry an error of up to 30\%. To correct the light-of-sight effect, the longitudinal magnetograms are then divided by cosine of the offset to the solar disk centre angle at each pixel~\citep[for details see][]{2001ApJ...555..448H}.


\par
In order to co-align the X-ray images with the chromospheric magnetograms, TRACE \mbox{171\,\AA\,} and 
\mbox{1700\,\AA\,} observations were used as intermediary. First we calibrated the TRACE data with \textit{trace\_prep.pro} 
which automatically aligns the observations of the two TRACE channels. We then co-aligned the X-ray images with 
the \mbox{TRACE 171~\AA} images and the SOT magnetograms with the \mbox{TRACE 1700~\AA} data. EIS \mbox{He\,{\sc ii}\,256.32~\AA\,} 
and \mbox{Fe\,{\sc xii}\,195.12~\AA\,} observations were also used to check the goodness of the 
co-alignment. 

\subsection{Analysis methods}
\label{subsect:method}

After the data calibration and alignment, {the brightening events identified  in paper II (in X-ray data) were overlayed on the SOT magnetograms.  Further, a  visual inspection was made to unsure which polarities are involved in an individual event. This was especially needed when a few events occurred close to each other.} 
In Fig.\,\ref{fig1} we show the X-ray images with each analysed event numbered and shown in dashed-line-boxed regions. The X-ray images in this figure were generated
by taking for each pixel the highest value obtained during the time intervals given in Table\,\ref{tab:obs}. We identified 22 events 
in the coronal hole datasets and six events in the quiet-Sun datasets as shown in Fig.\,\ref{fig1}.

\par
To follow the evolving magnetic features, the feature tracking software, SWAMIS \citep[The SouthWest Automatic 
Magnetic Identification Suite,][]{2007ApJ...666..576D, 2008ApJ...674..520L, 2010ApJ...720.1405L} was used in the present 
study. SWAMIS can be freely downloaded from \url{http://www.boulder.swri.edu/swamis}. The tracking procedures include feature 
detection, feature identification, feature association, tabulation and event classification. The details on the procedures 
and a user guide of the software can be found on the website. Gaps in the datasets can seriously mislead the tracking program, 
therefore, a linear interpolation was applied to fill these gaps. In SWAMIS, a low and a high threshold have to be assigned 
for the feature detection. For each dataset, five sigma of the magnetic field strength of all pixels from each magnetogram 
were used as the high threshold. In order to detect weak magnetic elements, we randomly selected 100 pixels in each data set which we determined 
as `weak'. The average magnetic strength of these pixels was determined to be \mbox{25~Mx~cm$^{-2}$}. Thus we used 
\mbox{25~Mx~cm$^{-2}$} as the low threshold. To detect a flux concentration (named hereafter `magnetic element'), 
a method called `downhill' was used in the feature identification. The `downhill' method determines first a local maximum 
of magnetic flux density and then expands down towards zero gradient of flux density which is defined as the edge of one 
magnetic element and its size. The magnetic element is then followed until it disappears or merges with another element, 
which determines its lifetime. The minimum size and lifetime of elements to be detected was set to four pixels 
(i.e. 0.64\arcsec$\times$0.64\arcsec) and three frames (i.e. 270 seconds for the CHs data and 135 seconds for the QS data). 
A new magnetic element can be born from splitting of the old magnetic element, or a flux emergence. SWAMIS defines a 
magnetic element that was born associated with a recently-born opposite magnetic element or with a growing opposite magnetic 
element as flux emergence. In our online material, we marked all magnetic elements on the magnetograms with symbols (flux 
emergence with squares, and the other new born magnetic elements with asterisks). However, these definitions do 
not always work as expected because of the very complex behaviour of magnetic features. Feature tracking is a very complex 
procedure and no automatic program can deal with it precisely~(Parnell, C. E., private communication), 
therefore, it has to be complemented by visual analysis. Each event was tracked by mean of the naked eye by simultaneously viewing of X-ray image and the corresponding magnetogram. All important changes were noted and organised in tables (see Table\,A\ref{tab:0901} as an example).\footnote{All the movies, together with 
the notes are provided as online materials at \href{http://www.arm.ac.uk/highlights/2012/603/bpmag.html}{www.arm.ac.uk/highlights/2012/603/bpmag.html}.}
\begin{figure*}[!ht]
 \centering
 \includegraphics[width=16cm, trim=25mm 42mm -10mm 35mm,clip]{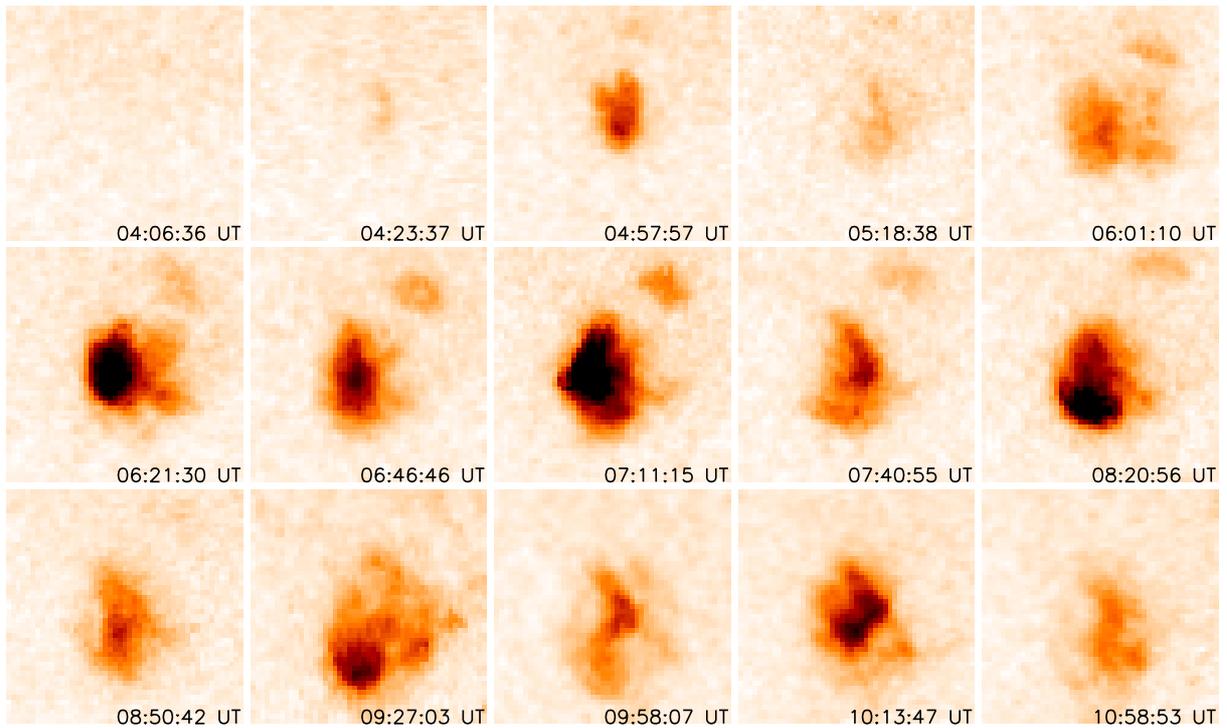}
 \caption{As Fig.\,\ref{fig6}, but for a CH bright point observed on November 12 (marked as No. 2 
 event in Fig.\,\ref{fig1}). The field-of-view has a size of 50\arcsec$\times$50\arcsec. An animated version is available online, c.f. Fig.\,A\ref{fig21}.}
 \label{fig9}
\end{figure*}

\begin{figure*}[!ht]
 \centering
 \includegraphics[width=16cm, trim=0mm -45mm 15mm 60mm,clip]{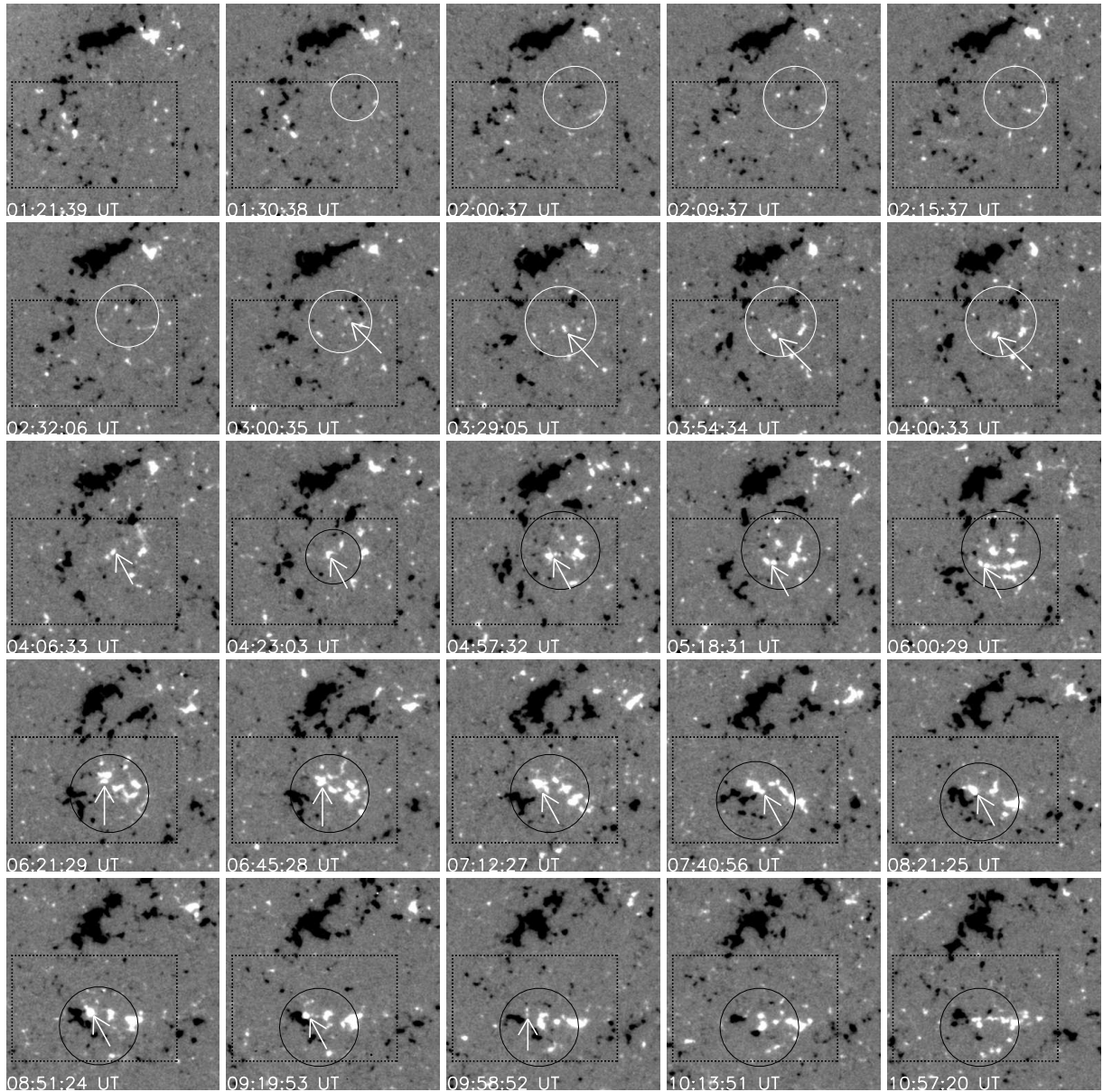}
 \caption{As Fig.\,\ref{fig7}, but for the bright point shown in Fig.\,\ref{fig9}. The magnetograms are scaled from --70~Mx~cm$^{-2}$to +70~Mx~cm$^{-2}$. The first two rows show the magnetic field evolution 
 before the emergence of the brightening event. The white arrow points at an emerging magnetic element which later cancels. The 
 white circles emphasise a bipolar region created by magnetic flux emergence, while the black circles emphasise the area when and 
 where magnetic cancellation occurs. Dotted lines outline the area where the magnetic flux lightcurve (Fig.\,\ref{fig11}) was
 extracted from. An animated version is available as online material, c.f. Fig.\,A\ref{fig21}.}
 \label{fig10}
\end{figure*}

\section{Results}
\label{sect:res}
\subsection{Magnetic field distribution in the coronal hole and the quiet Sun}
\label{subsect:magdistr}

We present in Figs.\,\ref{fig2} and \ref{fig3} the full SOT field-of-view (FOV) images of the coronal hole and the quiet-Sun regions respectively. 
A visual inspection of Fig.\,\ref{fig2} reveals that the dominant flux in the coronal hole is of negative polarity which also forms the largest concentration of magnetic flux in terms of size. 
To compare the distribution of positive and negative fluxes in both regions, we calculated the average flux for each magnetic element (see in 
Sect.~2.2 the definition of a magnetic element) over its lifetime. In Fig.\,\ref{fig4} we show the derived flux distribution for the four 
days of observations. In the CH the number of magnetic elements with negative flux is more than four times higher with respect to 
the positive flux elements. Both distributions peak at the same flux strength which is around {\mbox{3--4~$\times~10^{16}$~Mx}}, 
compared with {\mbox{2~$\times~10^{16}$~Mx}} in the quiet Sun. In the quiet Sun the number of magnetic elements is the same for 
both the negative and the positive flux. The small flux imbalance, especially on January 13, is due to the limited field-of-view. 
The fractional distribution of pixels with negative magnetic field can be found in Fig.\,\ref{fig5}. Again, the prevalence of a 
single polarity in the CH is very clear. The CH magnetic fields reach up to 600~Mx~cm$^{-2}$ while the quiet Sun contains magnetic field 
concentrations only up to 400~Mx~cm$^{-2}$. Almost 100\% of the highest values of the magnetic field in the coronal hole are found in pixels occupied by the dominant polarity. It should also be noted that the fractional distribution of positive and negative 
polarities in the coronal hole did not change over a three day period which separates the two coronal hole datasets. 
A flux stronger than 100~Mx~cm$^{-2}$ remains imbalanced in the quiet-Sun regions because of the small field-of-views.
 
\subsection{X-ray intensity variations and magnetic field evolution of brightening events in coronal holes}
\label{subsect:CH}
With SWAMIS, we were able to follow the magnetic elements from their birth to death, studying their behaviour including emergence, 
convergence, splitting, mergence and cancellation. All events in the coronal hole resulted from the interaction of bipolar magnetic 
regions. From 22 brightening events, seven were seen in the X-ray images {at} the start of the observations while 12 appeared 
during the observations but their corresponding bipolar regions were already present {at} the beginning of the observations. 
Three events were formed during the observations together with the formation of their corresponding bipolar region. In most of 
the cases (19 out of 22), one of the polarities forms a `stable' centre (which can be a single large polarity or a group of closely scattered 
small polarities). The stable polarity remains in the same place and in only a few cases when the 
 central polarity exists for more than a few hours, it may move a few arcseconds usually approaching the opposite polarity. The 
stable polarity tends to be the stronger polarity in the bipolar region. In eleven cases this is the negative polarity 
(i.e. dominant polarity in the coronal hole), in five cases it is the positive one, and in three cases it switches between 
polarities depending on which one is the strongest at a particular time remaining for hours each time. The remaining three cases are inconclusive.

\par
The evolution of the magnetic flux associated with a coronal transient brightening proceeds in a similar way for all events. All brightening 
events identified in the X-ray images are caused by magnetic flux emergence and a follow-up cancellation with the pre-existing 
and/or newly emerging magnetic flux. In some cases a magnetic flux (opposite to the stable magnetic element polarity) 
emerges close-by (a few arcsecs) from the stable polarity. The emergence is followed shortly after (a few tens of seconds) by 
cancellation. The emerging element can also surface at a distance away (20\arcsec\ for instance) from the stable center. The 
newly emerged polarity then starts moving towards the stable centre often becoming larger with time by merging with a flux of 
the same sign. Once the magnetic polarities get close to each other and start cancelling, a brightening will appear in the X-ray image. However, 
from the visual analysis we established that there are many more magnetic cancellation sites than X-ray brightening events. We 
 compared the magnetic flux (unsigned total) involved in magnetic cancellation while a brightening was observed in X-rays 
with the unsigned total magnetic flux when no X-ray brightening was present. We found that the magnetic flux involved in 
brightening events is twice larger than the cases without an X-ray response. We speculate that there is a threshold of the 
amount of magnetic flux involved in cancellation above which brightening would occur at X-ray temperatures. This subject, however, 
needs further detailed investigation which will be tackled in future work. Magnetic flux emergence is also important 
in maintaining the field strength of a magnetic element during its cancellation. Flux cancellation and emergence are 
often seen to happen simultaneously at the same location.
\begin{figure}[!ht]
 \centering
 \resizebox{\hsize}{!}{\includegraphics{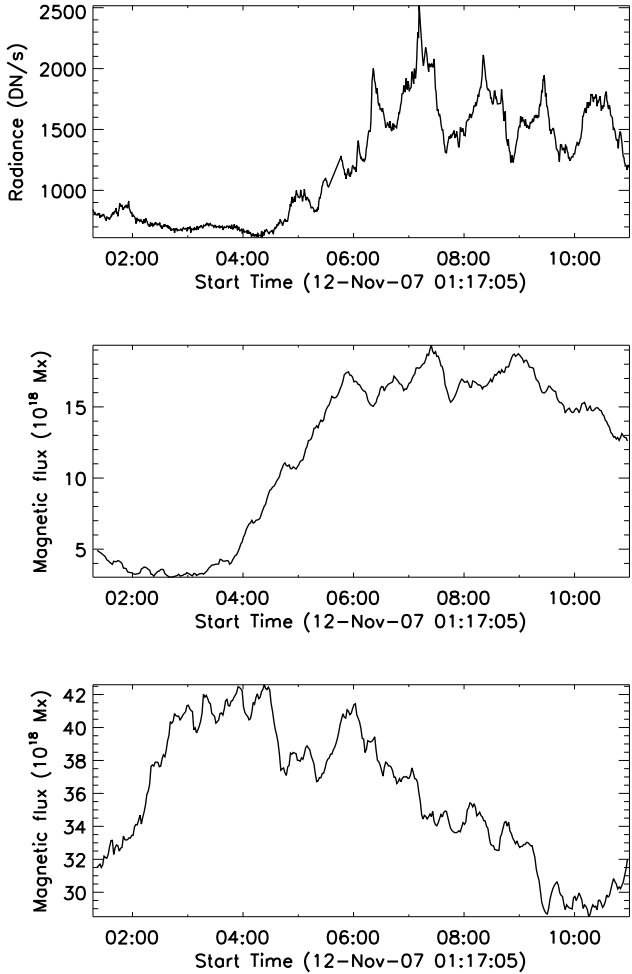}}
 \caption{As Fig.\,\ref{fig8}, lightcurves of a CH bright point that is marked as No. 2 event on November 12 in Fig.\,\ref{fig1}. 
 The X-ray lightcurve is extracted from the FOV shown in Fig.\,\ref{fig9} and the magnetic flux variations are extracted from 
 the region outlined by dotted lines in Fig.\,\ref{fig10}.}
 \label{fig11}
\end{figure}

\begin{figure}[!ht]
 \centering
 \resizebox{\hsize}{!}{\includegraphics{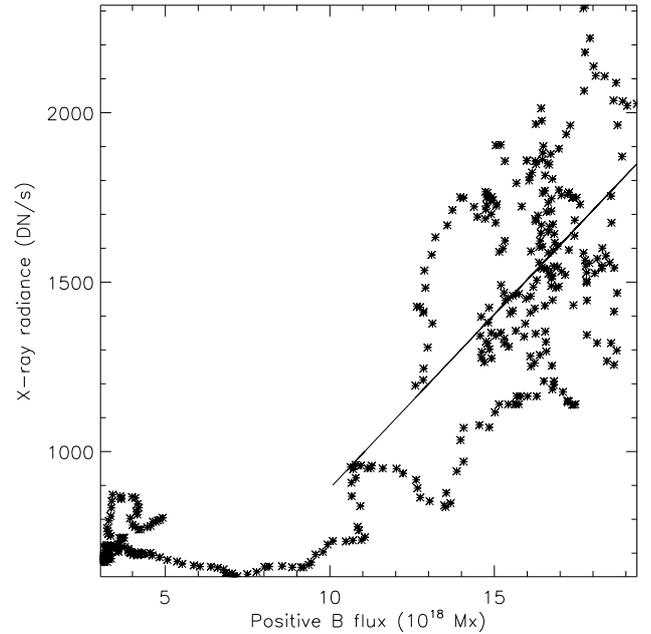}}
 \caption{Correlation analysis of the positive magnetic flux and the X-ray emission of event No. 2 on November 12.}
 \label{fig12}
\end{figure}

\par
The X-ray/magnetic-flux correlation analysis demonstrates a strong 
correspondence of the magnetic positive and/or negative fluxes with the {long-term (in the order of a few hours) changes in the X-ray emission (see the discussion 
 in the following subsection for more details). However, in some cases even changes as short as 1 hour can be observed (see the X-ray and positive flux variations in Fig.\,\ref{fig11} for example).}
The radiance variations of isolated bright points in the coronal hole (5 cases) correlated 
with {the magnetic flux changes in both polarities.} In 50\% of the cases (eleven) the radiance variations correlate ONLY 
with the positive flux, {i.e. the subordinate polarity in the CH}. In five out of 22 cases the correlation in time follows either 
one or the other polarity. In one case a very strong jet was followed in X-rays together with the emergence, 
evolution and disappearance of the magnetic flux associated with it.

\par
{Thanks to the very simple magnetic field configuration of the 
coronal hole (and in general any coronal hole) the evolution of each individual phenomenon can easily be analysed. In the following subsections, we give a detailed and further description of three examples (two CH bright points and one X-ray jet). Note that in the online material we give the same full description (figures, movies and tables) for each analysed event.}

\subsubsection{{Magnetic evolutions of two CH bright points}}
In Fig.\,\ref{fig6} we give an example of the temporal evolution of a CH bright point on November 9 (event No. 1) as 
seen in X-rays. The temporal evolution of the magnetic polarities associated with this event is presented in Fig.\,\ref{fig7}. 
The corresponding lightcurves in X-rays, positive and negative magnetic flux are shown in Fig.\,\ref{fig8}. A detailed description 
of the BP evolution is given in Table\,A\ref{tab:0901}. The intensity variation of the BP is apparent with peaks of the X-ray 
emission at 06:44~UT, 07:07~UT, 07:26~UT, 08:17~UT, 09:44~UT, 10:10~UT, 10:43~UT and 14:52~UT, with an overall periodicity of 
$\sim$20~mins. The shape and projected size of the BP change in time as well. We give an example of an emerging magnetic 
element (denoted by a white arrow in Fig.\,\ref{fig7}) which moves towards the pre-existed opposite polarity, causing 
magnetic cancellation. The magnetic 
element emerges at 07:21~UT and affects the event around 10:10~UT which results in a strong X-ray emission increase in this event 
(Fig.\,\ref{fig6} at 10:09:57~UT, and in Fig.\,\ref{fig8}, the highest peak in the lightcurve).

\par
Radiance oscillations in BPs have been the subject of several studies in the past (see Section\,\ref{sect:intro}). They were 
interpreted as the signature of propagating waves though suggestions were also made that these emission spikes can be due to 
repetitive magnetic reconnection. The lightcurves of the negative and positive magnetic fluxes derived here show similar 
fluctuations. The question is what is the origin of these small-scale variations of the magnetic flux? The lightcurves 
were smoothed with five data points to remove the short period variations. The five data points here correspond to 7.5~minutes 
(the cadence of our data is 1.5 minutes). A visual inspection of 
event No. 1 on November 9 and all other events suggests that the magnetic flux lightcurve variations are mostly due to 
emerging and cancelling magnetic fluxes.

\par
Event No.~2 on November 12 best describes the magnetic-flux/X-ray-emission correlation. In Fig.\,\ref{fig9} we show the X-ray image sequence and in Fig.\,\ref{fig10} 
the magnetogram sequence. The corresponding lightcurves are given in Fig.\,\ref{fig11}. The event first appeared in X-rays 
at 04:23~UT when a bipolar region was formed. The emergence of the positive magnetic elements in this bipolar region started at the 
beginning of the observations~(i.e. 01:21~UT). They emerged near a negative polarity and moved to another negative polarity to form 
the bipolar region. The negative polarity is the stable 
one and was formed from both pre-existing and newly emerging magnetic flux. The X-ray emission variations of the event have a period 
close to an hour. The positive flux strongly correlates with the X-ray emission (Fig.\,\ref{fig12}). The reoccurring positive 
magnetic flux cancellations are associated with an intensity variation and can be followed in the online material (Fig.\,A\ref{fig21}). 
An example of a magnetic polarity emergence and then cancellation is denoted by an arrow in Fig.\,\ref{fig10}. After emergence 
several small positive flux concentrations merge together forming larger flux concentrations. The cancellation with the negative 
magnetic elements starts at around 06:21~UT and the positive polarity disappears around 10:00~UT. This typical process can be 
summarised as: 
\textit{a single polarity emerging away from an opposite polarity magnetic element $\rightarrow$ becomes stronger and moves away 
from the emerging site $\rightarrow$ divides into a few elements, each of them becomes stronger and/or merges with other elements 
of the same sign$\rightarrow$ moves towards the opposite sign flux and cancels with it.}
\begin{figure*}[!ht]
 \centering
 \includegraphics[width=15cm, trim=-11mm -85mm 38mm 49mm,clip]{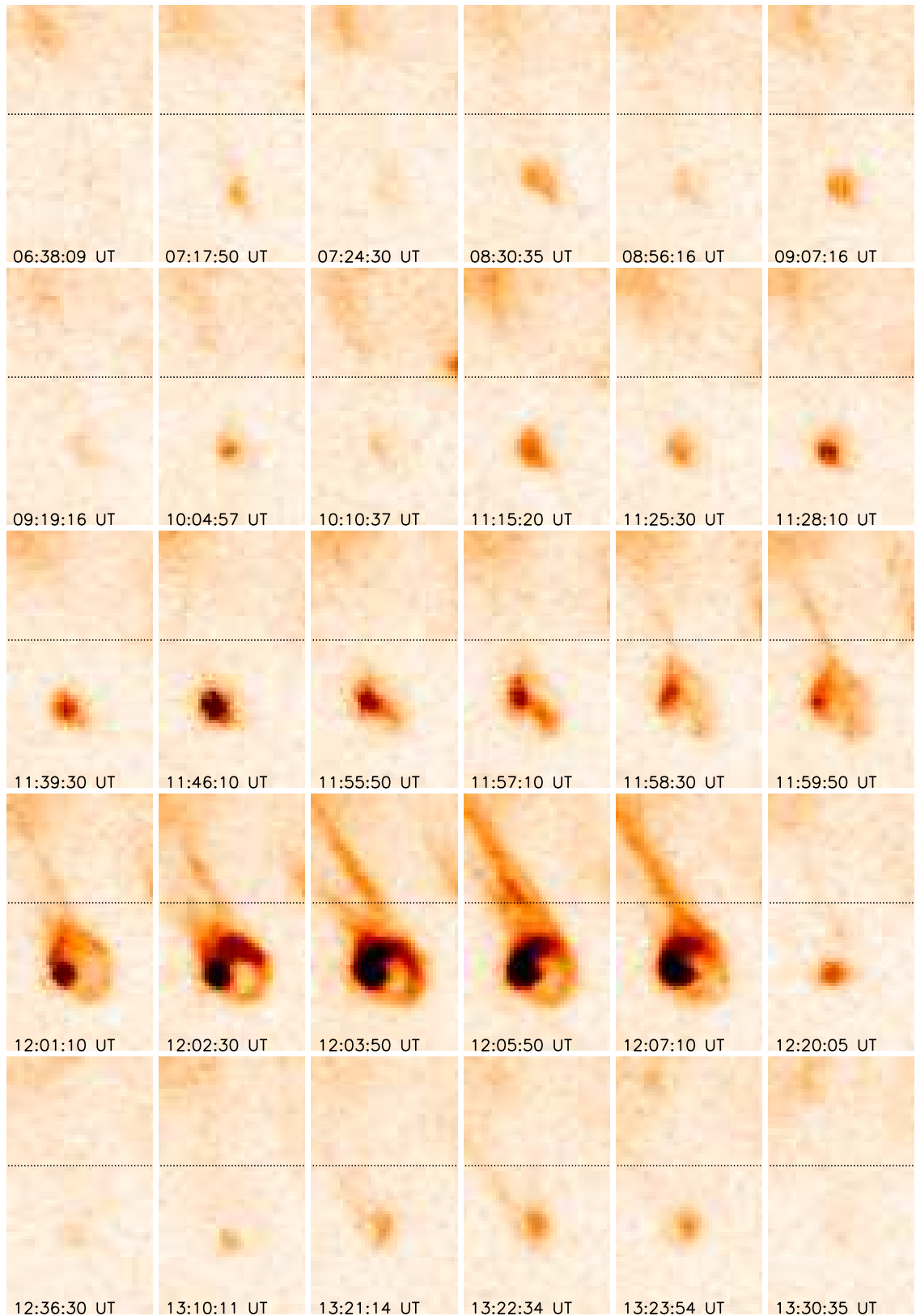}
 \caption{As Fig.\,\ref{fig6}, but for an X-ray jet in a CH on November 9 (marked as No. 3 event in 
 Fig.\,\ref{fig1}). The field-of-view has a size of 43\arcsec$\times$75\arcsec. The corresponding magnetograms of the area 
 below the black dotted line are shown in Fig.\,\ref{fig14}. An animated version is available online, c.f. Fig.\,A\ref{fig22}.}
 \label{fig13}
\end{figure*}
\begin{figure*}[!ht]
 \centering
 \includegraphics[width=15cm, trim=-11mm -85mm -7mm 100mm,clip]{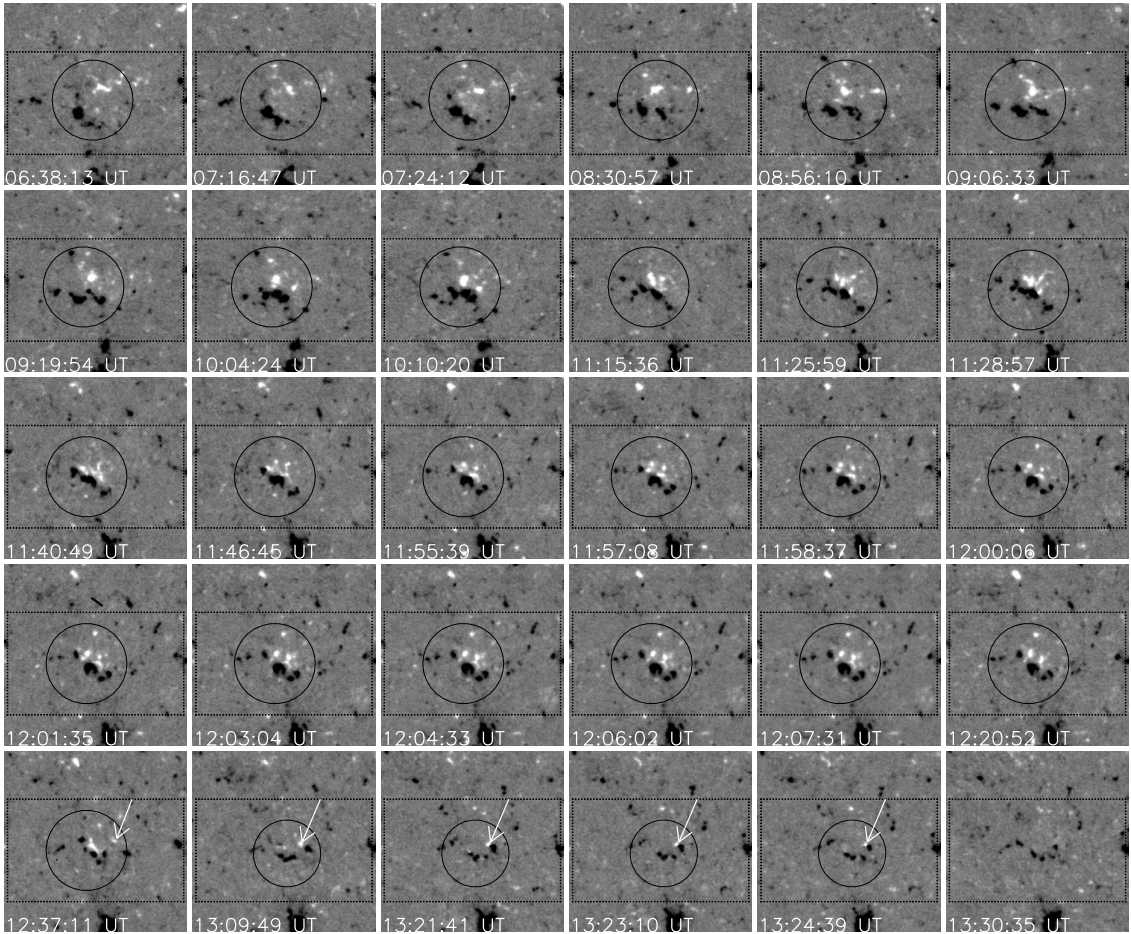}
 \caption{Longitudinal magnetic field images of the X-ray jet shown in Fig.\,\ref{fig13}. 
 The field-of-view corresponds to the area below the dotted lines in the X-ray images and has a size of 43\arcsec$\times$43\arcsec. 
 The magnetograms  are scaled from $-70$~Mx~cm$^{-2}$to $+70$~Mx~cm$^{-2}$. The black circles indicate the magnetic cancellation 
 sites. The white arrow points at an emerging positive magnetic element. The black dotted lines outline the region 
 where the magnetic flux lightcurves were extracted from. An animated version is available online, c.f. Fig.\,A\ref{fig22}.}
 \label{fig14}
\end{figure*}

\subsubsection{Magnetic flux evolution of an X-ray jet}
\label{subsect:jet}
\begin{figure}[!ht]
 \centering
 \resizebox{\hsize}{!}{\includegraphics{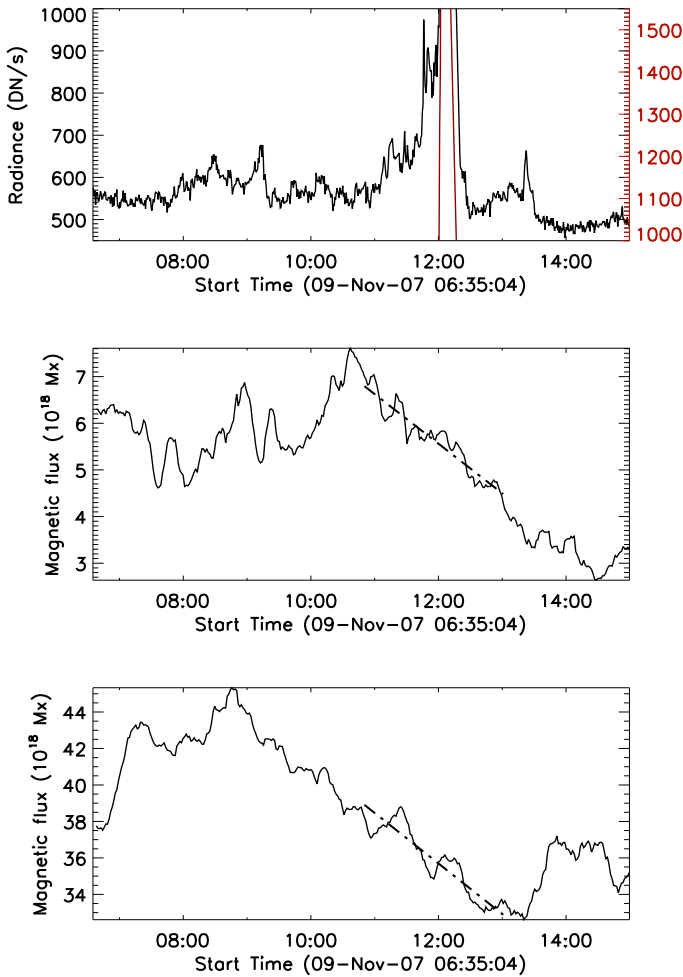}}
 \caption{Lightcurves of the X-ray jet marked as No. 3 event on  November 9  in Fig.\,\ref{fig1}. The X-ray lightcurve (top) is extracted from the FOV shown in 
 Fig.\,\ref{fig13} and the magnetic flux lightcurves (positive -- top and negative -- bottom) are extracted from the region outlined by black dotted lines in Fig.\,\ref{fig14}. 
 The time~(x) axis is shown in units of UT. {The peak of the X-ray lightcurve (top) is cut off and resume from below in red. The dash-doted-lines in the middle and bottom panels are linear fits of the corresponding magnetic flux from 10:50\,UT to 13:00\,UT.}}
 \label{fig15}
\end{figure}

In our observations, an X-ray jet was observed in the CH on 2007 November 9 (marked as No. 3 event in Fig.\,\ref{fig1}). The evolution of the jet in X-rays and its corresponding magnetic flux are shown in Fig.\,\ref{fig13} and \ref{fig14}. The X-ray and magnetic flux lightcurves are given in Fig.\,\ref{fig15}. At the beginning of the observations (06:38~UT) the region is very quiet, i.e. no X-ray emission variations are observed. Around 07:17~UT a bright point emerges, showing fluctuations in 
its X-ray emission until around 11:30~UT. At 11:44~UT a brightening in a few pixels is followed by growth of the bright point. 
At 11:55~UT, a loop structure, which could be the site projection of a sigmoid, is formed and starts rapidly expanding producing a jet 
around 11:58~UT. When the jet first forms, the brightening system (the loop and the jet) has an \mbox{inverted-Y-shaped}. 
Both the jet and the loop then become brighter. Around 12:01~UT, the loop appears O-shaped. Around 12:02~UT, the jet evolves into
two bright branches. The brightening system reaches peak emission around 12:05~UT, and then becomes weaker, returning to 
a single branch again around 12:07~UT. The jet disappears fully before 12:20~UT. Because of a data-gap between 12:08~UT 
and 12:20~UT, we could not follow the event during its weakening stage. A small bright point is still present when the 
observations resume at 12:20~UT. At 12:36~UT the bright point cannot be seen in X-rays. It flares again at 13:10~UT, and 
produces a small short-lived (about two minutes) jet around 13:21~UT. Around 13:24~UT, a small bright point remains  
but is weak. The brightening system completely disappears in X-rays after 13:30~UT.

\par
Fig.\,\ref{fig15} shows the variation of the magnetic flux in the footpoints of the jet. As often seen in bright points, the 
magnetic flux at the footpoints of the jet clearly decreases after its eruption. Fig.\,\ref{fig14} displays the longitudinal 
magnetograms evolution of the region. At 06:38~UT, a bipolar region is already present. Around 07:17~UT when the bright point 
emerges, positive elements are moving closer to the negative elements and some weak elements disappear due to cancellation. 
During the bright point intensity fluctuations, the positive elements continue moving towards the negative ones. The cancellation 
of the negative flux shown in the boxed region in Fig.\,\ref{fig14} starts shortly before 09:00~UT. After 09:00~UT the 
positive flux has also started to cancel, though additional positive flux emergence takes place before 10:00~UT. It is 
followed by a rapid flux cancellation starting at 10:40~UT. Around 11:40~UT, positive elements are touching the negative ones. At this time the negative fragments are closely grouped together. Around 11:55~UT when the BP evolves into a sigmoid-like 
feature, a positive element inserts into the space between the two separate negative ones. When the jet starts (at around 11:58~UT), 
the magnetogram shows that the positive element between two negative ones has become stronger and then starts to decrease. The 
positive element is very small and weak at 12:20~UT, it disappears at 12:37~UT. After the major jet, both negative and positive 
elements were more scattered and weaker, but the magnetic cancellation in the region continues. At around 12:22~UT, a 
small positive element emerges, and it is highlighted by a white arrow after 12:37~UT in Fig.\,\ref{fig14}. The emerged element 
becomes stronger and moves towards the negative one. It touches the negative element and becomes smaller at around 13:21~UT 
when the small jet is formed. It disappears at the same time as the brightening system seen in X-rays fully fades away. After 
13:30~UT, there are no positive elements visible while the negative elements in the region are much smaller than that at 
the beginning of the observation because of the cancellation.

\subsubsection{Magnetic flux cancellation rate of a BP and an X-ray jet}
\label{subsect:rate}
\par
We define the magnetic cancellation rate as $\frac{\Delta F}{\Delta T}$, where $\Delta F$ is the decrease of magnetic 
flux and $\Delta T$ is the time period of the decrease. In many cases the complex nature of the cancellation process does 
not allow us to calculate the cancellation rate. Fortunately, two events had a well isolated magnetic 
configuration which permitted such an estimation. This is the decaying stage of event No. 1~(bright point) on 9 
November (see Fig.\,\ref{fig6}) which was observed from 10:00~UT to 14:00~UT. During this time the magnetic flux lightcurves
(see Fig.\,\ref{fig8}) show a decrease after the largest flaring (see the peak in the X-ray light 
curve displayed in red). The X-ray jet discussed in Section\,\ref{subsect:jet} also has a clear magnetic 
structure (see Fig.\,\ref{fig14}) and shows a very clear decrease of the magnetic flux lightcurves around and after the 
eruption (see Fig.\,\ref{fig15} around 12:00~UT). {In Figs.\,\ref{fig8} and \ref{fig15}, we overplotted the corresponding linear fits on the corresponding unsigned magnetic flux of these two events.} We obtained the magnetic cancellation rates from the linear fit 
as $1.9\times10^{14}$~Mx/s~(positive polarity) and $6.5\times10^{14}$~Mx/s~(negative polarity) for the bright point, 
and $3.0\times10^{14}$~Mx/s~(positive polarity) and $7.9\times10^{14}$~Mx/s~(negative polarity) for the X-ray jet. {The flux cancellation rate of the positive polarity of the X-ray jet is about 60\% higher than that of the bright point. The magnetic flux cancellation rate of the negative polarities~(dominant one in this CH) of the X-ray jet is only 20\% higher with respect to the BP.}

\par
{As discussed previously, all brightening events are associated with magnetic cancellation. We presume that reconnection of higher loops has produced short loops which consequently submerged and this is possibly what we see at photospheric level as magnetic flux cancellation. We can only speculate that the higher rate of magnetic cancellation during the jet is a signature of a bursty physical process, e.g. magnetic reconnection, and the X-ray jet formation is the signature of this process. The jet observation presented here
is a single but very good example (unique concerning data cadence and resolution) of magnetic field evolution of an X-ray jet, however, drawing conclusions how all jets exactly evolve will be premature. }

\subsection{X-ray intensity variations and magnetic field evolution of brightening events in the quiet Sun}
Events in the quiet-Sun regions studied in the present study are similar to those in the coronal holes, e.g. they are all associated with bipolar 
regions, magnetic cancellation is found to correlate with their behaviour in X-rays. From six events, three emerged before the beginning of the observations, and three emerged during the observing periods but their related bipolar regions can be tracked 
from the start of the dataset. In the quiet-Sun region magnetic elements seed the whole region very densely and by moving 
frequently interact with each other in comparison to coronal holes. However, for the brightening events, a stable group of 
polarities is still found in three events. {Brightening events in the quiet-Sun region also show fluctuation in their X-ray emission, and repeated magnetic 
cancellation is found to be associated with it. We presently studied only six events in the quiet-Sun regions. Due to the small sample, we can not make any general conclusion on quiet-Sun events. Observations of SDO/AIA/HMI and forthcoming IRIS will provide an opportunity to include this subject in future study.
}

\section{Discussion and conclusion}
\label{sect:concl}
It is well known already for several decades that coronal holes form in unipolar photospheric magnetic fields. Thanks to 
the high-resolution longitudinal magnetic field observations provided by SOT/Hinode, we were able to follow the magnetic 
field evolution of quiet-Sun and coronal-hole regions at high cadence and small scale. We investigated the role of small-scale 
transients in the evolution of the magnetic field in an equatorial coronal hole. We used the magnetic feature tracking procedure 
SWAMIS to follow magnetic element emergence, movement, mergence, coalescence, cancellation etc. In addition, a visual analysis 
of each individual phenomenon simultaneously recorded by XRT and SOT was carried out. 

\par
We found that in the coronal hole the number of magnetic elements of the dominant polarity (in the present case, this is the negative polarity) 
is four times higher than the non-dominant one. The magnetic field concentrations in the coronal hole reach 600\,Mx~cm$^{-2}$ with respect to the quiet Sun 
where no concentrations with a field strength of more than 400~Mx~cm$^{-2}$ were detected. \citet{2010ApJ...719..131} found more kilo-Gauss magnetic 
field concentrations  in a polar coronal hole than in the quiet Sun. 
They explained the difference in the strength of the magnetic field of the two regions by the higher chance 
of collision, reconnection, magnetic energy loss and submergence of positive and negative elements in the quiet Sun which would 
easily weaken the field strength. We believe that additional factors such as a  different magnetic field diffusion and transportation in 
comparison to the quiet Sun as well as sub-photospheric processes may play a very important role. 
This will be a subject for future study.

\par
\citet{2006ApJ...649..464Z} analysed the magnetic flux distribution in a small coronal hole using high signal-to-noise level 
(2~G) magnetic observations from BBSO. They concluded that if one polarity dominates the network field, the opposite one will 
dominate the intranetwork field (i.e. weak field). {In this study, we also found that the dominant polarity in the coronal hole 
forms large concentrations of magnetic field while the opposite polarity is mostly dispersed in the form of weaker field concentrations. 
The magnetic flux distribution in the intranetwork region is not conclusive because its nature of weak field is a big 
challenge to the sensitivity and signal-to-noise of the instruments.} We noticed that the supergranulation configuration appear to have preserved 
its general shape during approximately nine hours of observations though the large concentrations in the network (the dominant 
polarity which sustains the coronal hole open magnetic field) did evolve and/or were slightly displaced, and their strength either increased or decreased. 
All changes were caused by the interaction of the network field~(mostly the dominant polarity) with the emerging opposite 
polarity magnetic field. We observe in 19 of 22 events in the coronal hole a single stable polarity which in eleven cases is the dominant polarity. That strongly suggests that the formation of small-scale transients is due to the interaction of pre-existing (often long-lived) magnetic flux with newly emerging opposite polarity flux.

\par
Our results show that all brightening events are associated with 
bipolar regions. A bipolar region is not formed by only two simple opposite magnetic polarities but each polarity is organised in many magnetic elements. From KPNO observations, \citet{1977SoPh...53..111G} found that bipolar regions did not exist in newly emerged and decayed X-ray bright 
points. Here, some of the studied events were followed from their emergence, through their entire lifetime until 
their full disappearance. Bipolar regions were found in the footpoints of the events before the event appearance at X-rays and 
were still present (although weak) after the decay of the feature in X-rays. This shows that
transient brightenings can maintain high temperature plasma only during part of their lifetime. During the emerging and 
decaying stages of X-ray bright points, the magnetic flux was found to be relatively weak which is possibly the reason for it not to 
have been detected in the KPNO observations of \citet{1977SoPh...53..111G}.

\par
Thanks to the open magnetic flux in coronal holes, plasma at high speed, temperature and density is 
ejected to far away distances as a result of magnetic reconnection forming 
X-ray/EUV jets (see paper III for more discussion). In the present study, the magnetic flux evolution associated with an X-ray jet was presented 
in unprecedented detail. Similar to X-ray bright points, magnetic cancellation 
was observed during the eruption. In comparison to a bright point, however, the rate of magnetic cancellation occurring in the footpoints of the X-ray jet was twice higher. \citet{1998SoPh..178..379S} analysed the magnetic flux evolution of 25 X-ray jets observed by the Soft X-ray telescope onboard \textit{Yohkoh} using full-disk KPNO longitudinal magnetograms. They found that 72\% of the jets occurred in mixed polarity region. We believe that all brightening events including jets are due to the interaction of bipolar regions. It should be noted that the SOT observations have higher sensitivity and more than ten times higher spatial resolution with respect to the KPNO data. That allowed us to detect much smaller and weaker magnetic flux and as noted earlier, weak magnetic fields are often responsible for the formation of various small-scale transients. More studies based on high-resolution observations of X-ray jet magnetic fields are needed to fully understand their formation patterns.

\par
Following reconnection, the newly formed smaller loops could eventually submerge which is observed as cancellation of magnetic dipoles \citep{1999SoPh..190...35H}. Much more magnetic cancellation sites than X-ray 
brightening events were observed in the present paper, i.e. not all magnetic cancellations relate to X-ray brightenings. This result is in agreement 
with \citet{1977SoPh...53..111G}, who found that not all active magnetic regions (namely ephemeral active regions) coincide with X-ray bright points.   In order to observe brightening in X-rays, plasma has to be heated to X-ray temperatures, which means that only reconnection sites which release sufficient energy can have an X-ray response.

\par
In paper III, we have found that transient brightening events have a response over a wide temperature range from the chromosphere 
to the corona. This result, however, was obtained using spectroscopic EIS and SUMER slit observations, which means that we 
had a very limited view and cadence to investigate how cancellation of magnetic flux with different strength is related to 
chromospheric, transition region or/and coronal emission. The multi-wavelength observations from 
AIA~\citep[The Atmospheric Imaging Assembly,][]{aia2010} and 
HMI~\citep[The Helioseismic and Magnetic Imager,][]{hmi2010} on board \textit{SDO}~\citep[The Solar Dynamics Observatory,][]{sdo2010} 
 provide an unique opportunity to further investigate this. We should be able to find observational signatures on what is the relation between photospheric magnetic flux cancellation and magnetic reconnection, and 
whether magnetic cancellation (submergence or other processes) is always related to energy release into the solar atmosphere. 

\citet{2005ApJ...626..563F} modelled the transport of open magnetic fields in the Sun. He suggested that 
the coronal holes are formed because of a local minimum of dipolar magnetic flux emergence. 
This idea was later supported from analysing observations from \textit{SOHO}/MDI~\citep{2006ApJ...641L..65A} and 
BBSO~\citep{2006ApJ...649..464Z}, where they found dipolar flux emergence rates in coronal holes lower than in 
quiet-Sun regions. Our study demonstrates that the magnetic flux in coronal holes 
is continuously `recycled' through magnetic reconnection which is responsible for the formation of numerous 
small-scale transient events. The open magnetic flux forming the coronal-hole phenomenon is largely involved 
in these transient features. The question on whether this open flux is transported as a result of the 
formation and evolution of these transient events, however, still remains open. Additional analysis is needed 
comprising of field extrapolations from high-cadence and high resolution magnetic field observations, together 
with MHD modelling of magnetic reconnection at coronal holes boundaries and comparison with high-resolution 
imaging and spectroscopic information. Data from AIA and HMI on SDO may be a suitable first step towards such an analysis.

\begin{acknowledgements}
We gratefully acknowledge the anonymous referee for the very important corrections and constructive suggestions. We thank Miss Kamalam Vanninathan for her careful and critical reading of the manuscript. Research at Armagh Observatory is grant-aided by the N.~Ireland Department of Culture, Arts and Leisure (DCAL). ZH thanks DCAL for the PhD studentship. We also thank STFC for support via grants ST/F001843/1 and PP/E002242/1. The research leading to these results has received funding from the European Commission's Seventh Framework Programme (FP7/2007-2013) under the grant agreement eHeroes (project n$^{\circ}$284461, www.eheroes.eu). Hinode is a Japanese mission developed and 
launched by ISAS/JAXA, with NAOJ as domestic partner, and NASA and STFC (UK) as international partners. It is operated by these agencies in co-operation with ESA and NSC (Norway). SWAMIS was written by Craig DeForest and Derek Lamb at the Southwest Research Institute Department of Space Studies in Boulder, Colorado. MSM thanks ISSI Bern for the support of the team ``Magnetic Flux Emergence in the Solar Atmosphere''.
\end{acknowledgements}
\bibliography{references}

\Online
\appendix
\begin{table*}[!h]
\centering
\caption{A timeline on the X-ray and magnetic field evolution of a CH BP (marked as No. 1 event in Fig.\,\ref{fig1}) on 2007 Nov. 9.}
\begin{tabular}{llll}
\hline
Time (UT) & X-ray BP behaviour & Magnetic fragment behaviour & Notes\\
\hline
06:38 & the BP is already present & A simple bipolar region & Beginning of the dataset\\
	 \hline
06:45 & BP reaches an intensity peak & A positive fragment moves towards a negative &\\
 & & fragment and magnetic cancellation takes place & \\
 \hline
06:57 & BP becomes weaker & The cancelling positive fragment & This scenario repeats \\
 & & disappears & around every 25 minutes on average. The \\
 & &  & fluctuation always relates to magnetic \\
	 & &  & cancellation and the cancellation site is \\
	 &  &   & always near to major positive polarities, \\
	 &  &   & but sometimes only negative fragments \\
	 &  &   & move to the site and at other times both \\
	 &  &   & polarities move and join together.\\
	 \hline
07:52 & 		 & Negative fragments start to  &\\
 &			 &emerge at about 10\arcsec\ away  &\\
	 &			 & from the positive ones, then  &\\
	 & 			 &start moving towards the positive  &\\
	 &			 & polarities.    & \\
	 \hline
08:22 & A small BP emerges at the & A few weak negative fragments emerge west &\\
 & west side.  & of the domain positive fragments & \\
	 &   & and cancellation occurs between these & \\
	 &   & two polarities. & \\
	 \hline
08:43 & The small BP disappears. & One positive fragment at the west part &\\
 &  & disappeared after cancellation with newly &\\
	 &  & emerging negative fragments. &\\
	 \hline
09:45 & The main BP reaches its second & The emerged negative fragments at 07:52 move & A few negative fragments disappear after\\
 & highest intensity peak. & closer to positive polarities and cancellation occurs. & this time and that leads to\\
 &&&the BP fading again\\
 \hline
10:10 & The main BP reaches its & The rest of the emerged negative fragments keep & Again it fades after the disappearance of a \\
 & highest intensity peak. & moving towards the positive polarities and cancel. & few negative fragments, however existing \\
	 &  &    & negative fragments keep cancelling and\\
	 &  &    & totally disappear at 11:25~UT creating\\ 
	 &  &    & a few peaks in the lightcurve of the BP.\\
	 \hline
10:36 & A new BP emerges southwest & Another negative polarity to the southwest & Cancellation and disappearance of this \\
 & of the main one. & move towards the positive polarities. & group of negative polarities take control\\
	 &  &   & of the behaviour of the new born BP.\\
	 \hline
12:31 & BP is almost fully gone. & Both positive polarities and the southwest & The main BP is very weak but still visible \\
 & & negative polarities move towards each other, & When some other negative fragments \\
	 &  & however, no cancellation was visible. & move closer to the positive fragments, \\
	 &  &   & the BP flames up again until the end of the \\
	 &  &   & dataset, the positive polarities still \\
	 &  &   & exist, and the BP repeats its intensity\\
	 &  &   & fluctuation.\\
\hline
\end{tabular}
\label{tab:0901}
\end{table*}

\begin{figure*}[!ht]
\centering
\includegraphics[width=15cm]{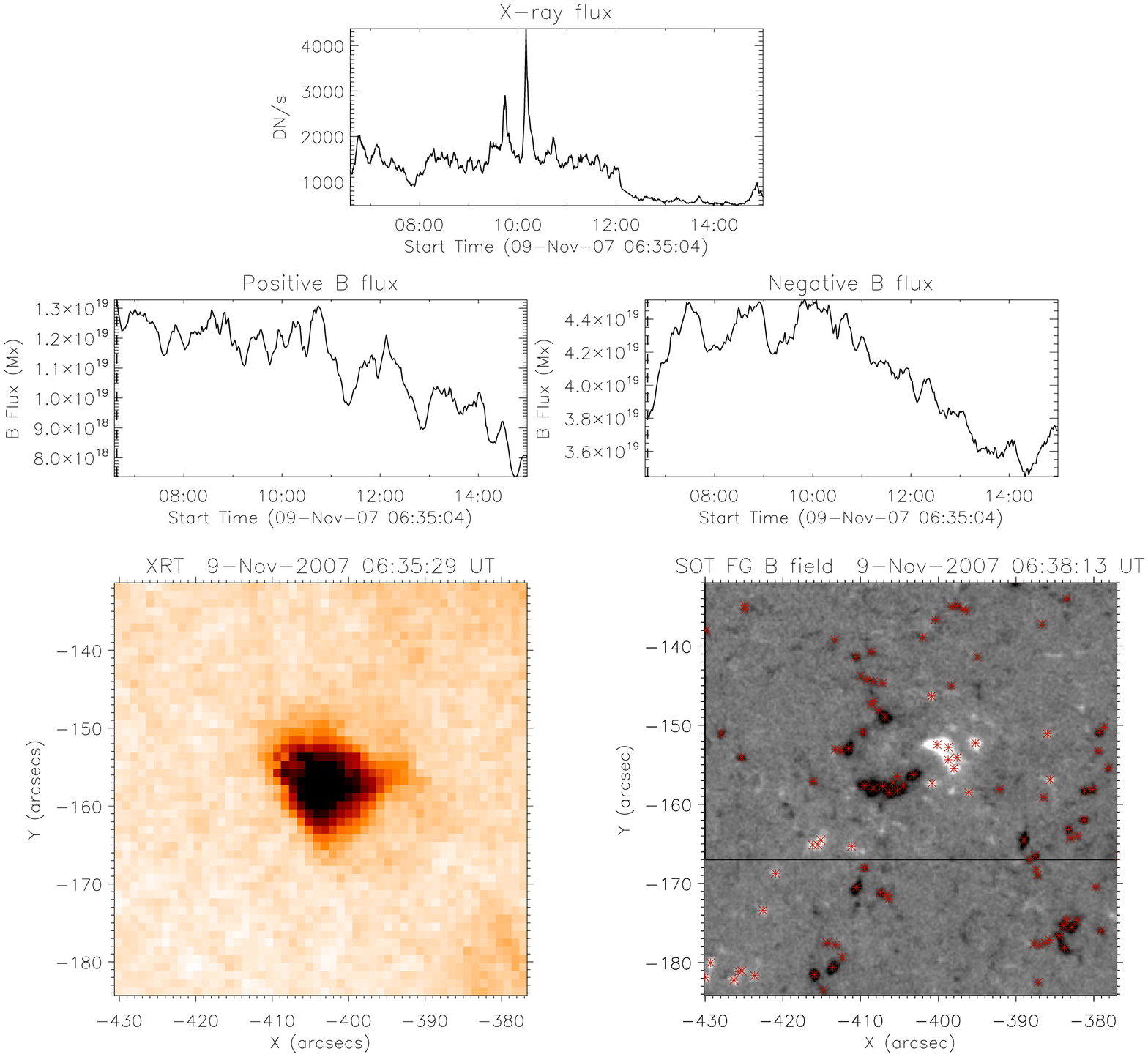}
\caption{An animation showing evolution of CH event No. 1 on 2007 November 9~(Figs.\,\ref{fig6}, and \ref{fig7}). The top panel is the 
X-ray lightcurve; the middle panel is the lightcurves of positive~(left) and negative~(right) longitudinal magnetic flux; the bottom 
panel shows the evolution of the event seen in X-ray images~(left) and SOT FG longitudinal magnetograms~(right). The time axis in the lightcurve 
plots is showen in UT. During the animation, the dashed line in the lightcurve indicates the given time of the X-ray image and longitudinal 
magnetogram~(bottom panel). Solid lines in the magnetograms outline the region where the magnetic flux lightcurves were extracted. Symbols 
indicate magnetic elements tracked by SWAMIS. Asterisks are elements first detected at the given time. Cross symbols indicate 
magnetic elements tracked from the past. Square symbols indicate magnetic elements first detected and defined by SWAMIS as 
flux emergence. (Please note that in this figure only asterisks are showed because it is the first frame of the observation, and other symbols will present through the animation.)}
\label{fig20}
\end{figure*}

\begin{figure*}[!ht]
\centering
\includegraphics[width=15cm]{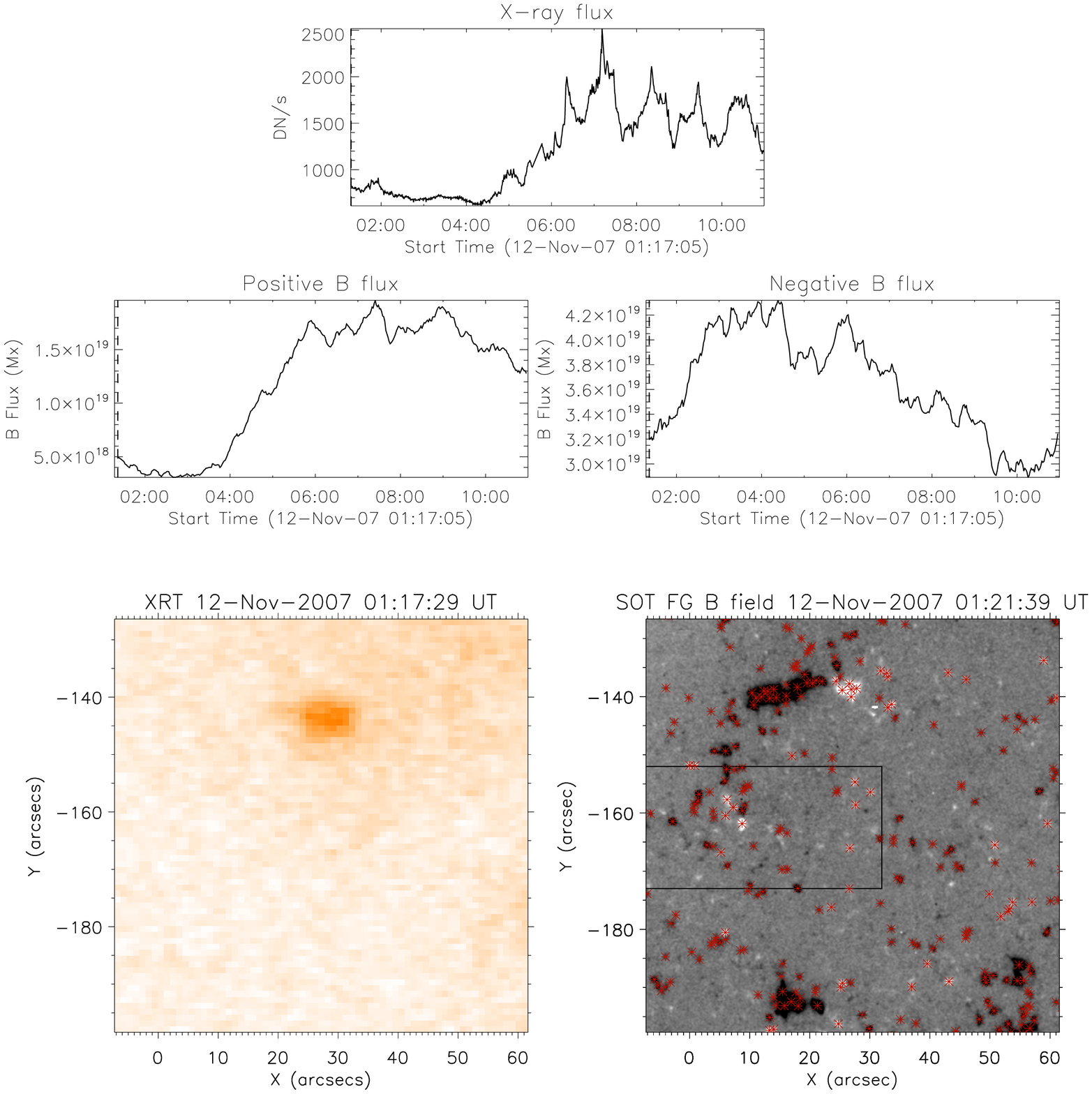}
\caption{As Fig.~A\ref{fig20}, but for CH brightening event No. 2 on 2007 November 12~(Figs.\,\ref{fig9}, and \ref{fig10}).}
\label{fig21}
\end{figure*}

\begin{figure*}[!ht]
\centering
\includegraphics[width=15cm]{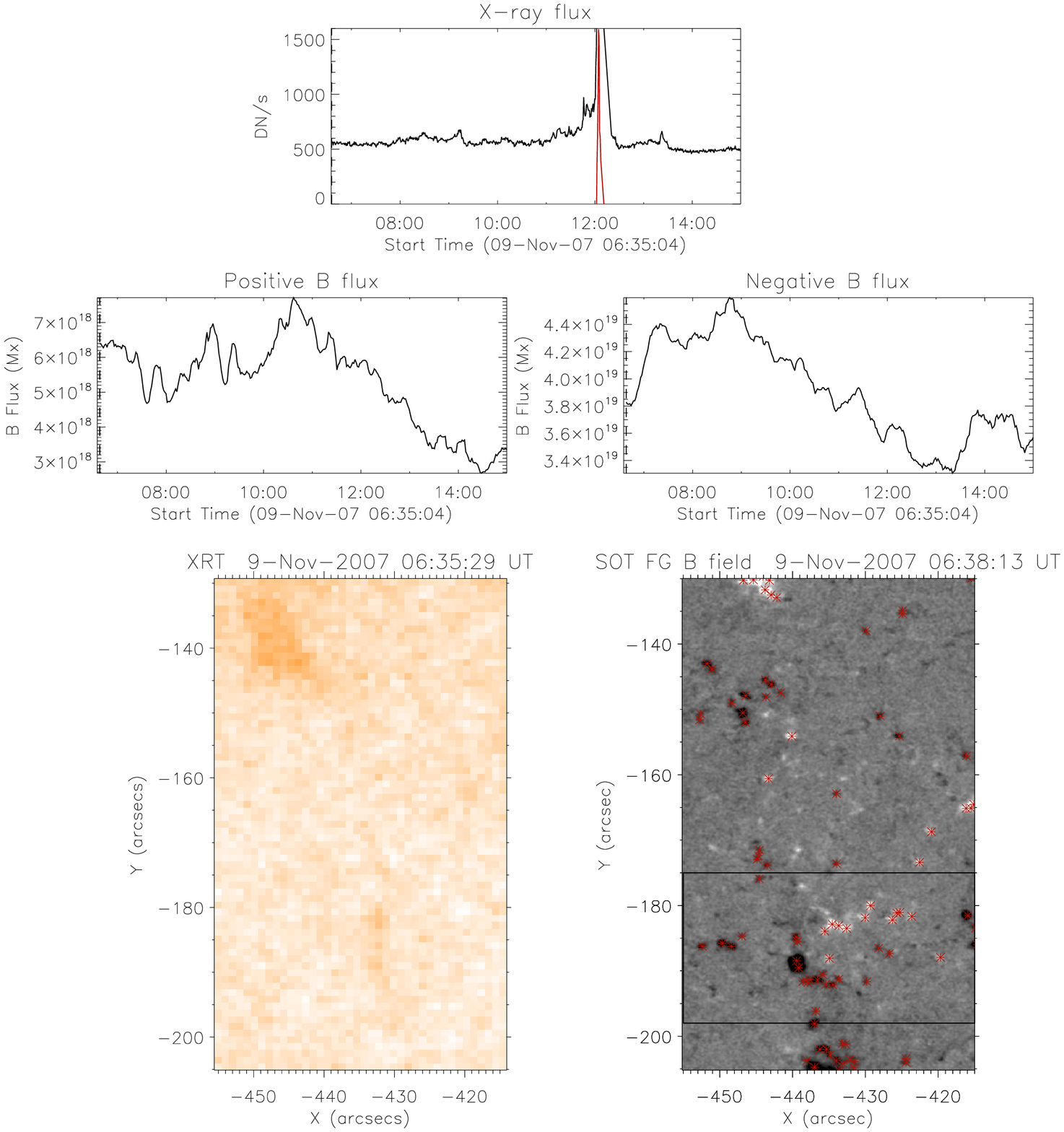}
\caption{As Fig.~A\ref{fig20}, but for the CH X-ray jet, No. 3 on 2007 November 9~(Figs.\,\ref{fig13}, and \ref{fig14}).}
\label{fig22}
\end{figure*}


\end{document}